\documentclass[prc,twocolumn,secnumroman,tightenlines,amssymb,aps,nobibnotes,
  superscriptaddress,showpacs,balancelastpage,floatfix,nofootinbib]{revtex4}
\usepackage{graphicx,colordvi}
\usepackage{multirow}
\usepackage{color}
\usepackage[utf8]{inputenc}
\usepackage{amsmath}

\begin{document}

\title{\bf Fourier-over-Spheroid shape parametrization applied to nuclear fission dynamics}
\author{K. Pomorski}\email{pomorski@kft.umcs.lublin.pl}
\affiliation{Uniwersytet Marii Curie Sk\l odowskiej, Katedra Fizyki
Teoretycznej, 20031 Lublin, Poland}
\author{B. Nerlo-Pomorska}
\affiliation{Uniwersytet Marii Curie Sk\l odowskiej, Katedra Fizyki
Teoretycznej, 20031 Lublin, Poland}
\author{C. Schmitt}
\affiliation{IPHC, University of Strasbourg, 67200 Strasbourg, France}
\author{Z. G. Xiao}\email{xiaozg@mail.tsinghua.edu.cn}
\affiliation{Department of Physics, Tsinghua University, Beijing 100084, China}
\author{Y. J. Chen}
\affiliation{China Institute of Atomic Energy, Beijing 102413, China}
\author{L. L. Liu}
\affiliation{China Institute of Atomic Energy, Beijing 102413, China}

\date{\today}
\begin{abstract}
\noindent

We propose a new, rapidly convergent, the so-called Fourier over Spheroid (FoS), shape parametrization to model fission of heavy nuclei. Four collective coordinates are used to characterize the shape of the fissioning system, being its elongation, left-right asymmetry, neck size, and non-axiality. The potential energy landscape is computed within the macroscopic-microscopic approach, on the top of which the multi-dimensional Langevin equation is solved to describe the dynamics. Charge equilibration at scission and de-excitation of the primary fragments after scission are further considered. The model gives access to a wide variety of observables, including fission fragments mass, charge, and kinetic energy yields, fragment mean N/Z and post-scission neutron multiplicities, and importantly, their correlations. The latter are crucial to unravel the complexity of the fission process. The parameters of the model were tuned to reproduce experimental observation from thermal neutron-induced fission of $^{235}$U, and next used to discuss the transition from the asymmetric to symmetric fission along the Fm isotopic chain.\\

\end{abstract}
\maketitle

  
\section{Introduction}

Fission is a dynamical process along which a nucleus progressively deforms (either spontaneously or triggered by an external perturbation) from an initial compact configuration until a point where it splits into two fragments. This evolution is an intricate puzzle, involving a complex re-arrangement of the many-body neutron and proton quantum systems. Intense effort is invested in fission studies since its discovery, both on the experimental and theoretical front, due to the impact for fundamental nuclear physics and astrophysics, as well as for a wide variety of societal applications.\\
Modeling fission, in general, implies four stages: (i) the definition of the initial conditions of the system, (ii) its dynamical evolution, and rearrangement in specific configurations of fragment pairs with corresponding probabilities, (iii) the (fast) prompt de-excitation of the excited fragments, and (iv) the (slow) decay towards $\beta$-stability of those fragments which are radioactive. The recent review by Schunck and Regnier \cite{SRe22} gives an excellent panorama on contemporary fission theories, and further details about foundations can be found in the textbook by Krappe and Pomorski \cite{KPo12}. In spontaneous and low-energy (mostly induced by neutrons) fission, the initial conditions are well defined. The radioactive decay of the fission products is well known also. To understand fission, the challenge thus mainly resides in the description of stages (ii) and (iii). These are not independent one from another: Stage (iii) critically depends on the properties ($N$, $Z$, excitation energy and angular momentum) of the (primary) fragments produced at scission at the end of stage (ii).\\
While experimental information was restricted to fission-fragment mass distributions with limited resolution for several decades \cite{ANS18}, recent developments give access to a widespread variety of observables, their correlations, and this with unprecedented resolution \cite{GHO18,ATJ20, CDF13, RCF19, martin:2021}. Such information is essential to unravel in an un-ambiguous way the intricacies of the fission process. It is obviously of primary importance for constraining theory, but it poses also a tremendous challenge, which is the requirement of modeling all aspects of the mechanism and their mutual interdependences.\\
\\
According to the complexity of the fission process, its description remains a challenge for theory, and various models have been proposed over the years. The last decade has seen the tremendous development of microscopic, self-consistent models. Unfortunately, quantitative description remains limited so far, and computing time makes systematic calculations impossible even at super computers \cite{SRe22}. Transport models within the macroscopic-microscopic approach have been established as a very good alternative. In this framework, the process is given by the solution of a classical equation of motion picturing the real-time evolution of the system on its potential energy landscape (PEL) under the influence of inertia, dissipation and fluctuations \cite{AGD86}. Systematic studies covering different regions of the nuclear chart are nowadays computationally tractable. Such widespread investigations are indispensable to converge towards a universal understanding of the process \cite{mahata:2022}.\\
Sophisticated macroscopic-microscopic models based on the solution of the multi-dimensional Langevin equation, or some variant of it, were developed during the last two decades \cite{SRe22,SJA16}. In these models, three main ingredients are required: a parametrization of the nuclear shape involving as few as possible deformation coordinates, a prescription for the potential energy of the nucleus, and a modelization for inertia and friction forces. Aritomo et al. {\cite{ACI14} and Usang et al. \cite{UII19} developed, respectively, a 3D and 4D dynamical model for explaining fragment mass and total kinetic energy (TKE) distributions in spontaneous and low-energy fission. Unfortunately, these models do not compute the post-scission de-excitation of the fragments. Furthermore, the hypothesis of unchanged charge density (UCD), {\it i.e.} the fragments have the same $N$/$Z$ ratio as the fissioning system, is assumed in the model of Aritomo et al. Finally, evaporation prior scission (so-called multi-chance fission) is not considered, making these codes un-suited for fissioning system initial excitation energy above 10 MeV or so \cite{hirose:2017}. The Brownian shape motion model by Randrup and M\"oller \cite{RMo13} is based on today highest-quality 5D potential energy landscapes. While its enhanced version by Albertsson et al. \cite{ACD21} adds the post-scission stage, similarly to the early code, the UCD assumption is made. M\"oller and Ichikawa \cite{MIc15} went beyond this hypothesis, treating independently neutrons and protons, what renders the model "6D". Unfortunately, this version is still to be combined to the post-scission stage of Ref. \cite{ACD21}. Furthermore, like for Refs. \cite{ACI14,UII19} the possibility of multi-chance fission is not implemented. In our previous works \cite{PNB15,SPN17}, we have developed an innovative nuclear shape parametrization, the Fourier parametrization, which demonstrated to gather within 4 collective coordinates the main features of the shapes relevant to fission. The new shape parametrization was succesfully used within the Born-Oppenheimer approximation \cite{PIN17} to describe fission fragment mass yields \cite{PDH20,PBK21}. We further implemented this parametrization (restricted to 3D), with a suited PEL prescription, and inertia and friction forces borrowed from classical mechanics, into a Langevin code. The latter showed able to reasonably describe fragment mass and TKE distributions from low-energy fission of typical actinides \cite{LCW21}. It was used also for predictions in the super-heavy element region \cite{KDN21}. The present work is a two-fold extension of these papers. First, we present an enhanced version of our shape parametrization, called the Fourier over Spheroid (FoS) \cite{PNB17,PNe23}. Second, we develop the previous Langevin code by proposing a method to compute (i) the fragments ($N$, $Z$) composition {\it i.e.} levelling off the UCD assumption, and (ii) their properties in terms of excitation energy and deformation at the instant of scission \footnote{At present, the angular momentum of the fragments is not treated in the model.}. This information is finally used as input in the extension of the code to the calculation of the post-scission stage. Altogether is demonstrated to offer a particularly fast and flexible way to compute a wide variety of observables. Comparison with experiment is made wherever possible for spontaneous and low-energy fission. Although not treated in this manuscript, work to account for multi-chance fission is in progress.

\section{Model}

In this section the various ingredients entering in the here-developed model are presented. Thermal neutron-induced fission of $^{235}$U is taken as an example to illustrate the main features of the theory and the variety of observables computed by the code. In section \ref{sec_fm} the model is applied to spontaneous fission of fermium.

\subsection{Nuclear shape parametrization}

The surface of the fissioning nucleus is described in the cylindrical coordinates $(\rho,\varphi,z)$ by the following formula \cite{PNe23}:  
\begin{equation}
\rho^2(z,\varphi)=\frac{R_0^2}{c}\,f\left(\frac{z-z_{\rm sh}}{z_0}\right)
{1-\eta^2\over 1+\eta^2+2\eta\cos(2\varphi)} ~,
\label{rhos}
\end{equation}
where $\rho(z,\varphi)$ is the distance from the $z$-axis to the surface. Function $f(u)$ defines the shape of the nucleus having half-length $c=1$:
\begin{equation}
  f(u)=1-u^2-\sum\limits_{k=1}^n \left\{a_{2k}\cos({k-1\over 2}\pi u)
             +a_{2k+1}\sin(k\pi u)\right\}~,
\label{fos}
\end{equation}
where $-1\leq u \leq 1$ and the expansion coefficients $a_i$ are treated as the the deformation parameters. The first two terms in $f(u)$ describe a sphere. The volume conservation condition implies $a_2=a_4/3-a_6/5+\dots$. The parameter $c$ determines the elongation of the nucleus keeping its volume fixed, while $a_3$ and $a_4$ describe the reflectional asymmetry and the neck size, respectively. The half-length is $z_0=cR_0$, where $R_0$ is the radius of a sphere with the same volume. The $z$-coordinate varies in the range $-z_0+z_{\rm sh}\leq z\leq z_0+ z_{\rm sh}$. The shift $z_{\rm sh} = -3/(4\pi)\,z_0\,(a_3-a_5/2+\dots)$ places the mass of the nucleus at the origin of the coordinate system. The parameter $\eta$ describes a possible elliptical, non-axial deformation of a nucleus.

The formula (\ref{rhos}) is entirely equivalent to those based on the Fourier expansion and described in Refs.~\cite{SPN17}. Here, the deviation from a sphere with radius $\rho=1$ is firstly expanded in the Fourier series, and subsequently, this deformed object of the length $2R_0$ is scaled to the elongation equal to $2cR_0$. The formula (\ref{rhos}) is more adapted to the calculation of the PEL of nuclei made on a mesh in the multi-dimensional deformation parameter ($c,a_3,a_4,...,a_n$) space since the range of variability of the $a_i$ coefficients does not depend on the elongation $c$. In addition, the mass ratio of the fragments, their relative distance, and the radius of the neck between them, measured in $z_0$ units, do not depend on the elongation of the nucleus. It is also worth noticing that for the reflection symmetric shapes, the geometrical scission points appear when $a_4=a_4^{\rm sc}={3\over 4}+{6\over 5}a_6\dots$ independently of the elongation $c$. Such properties of the present FoS shape parametrization make it very useful for all kinds of calculations related to nuclear fission.

The PELs of fissioning nuclei are obtained in the 4D space of deformation parameters ($c,a_3,a_4,\eta$) using the macro-micro model \cite{NTS69}. The macroscopic part of the energy is evaluated according to the Lublin-Strasbourg-Drop (LSD) formula \cite{PDu09}, while the microscopic energy corrections are calculated using the Yukawa-folded single-particle potential \cite{DPB16} and the Strutinsky shell correction method \cite{Str66,NTS69}. The pairing correlations are described using the BCS formalism \cite{BCS57} using an approximative projection on a good particle number\cite{GPo86,PPS89}. All parameters of the macro-micro model used in the present paper are the same as in Ref.~\cite{PDN22}.

A typical PEL of the $^{236}$U fissioning nucleus as an example  is shown in Fig.~\ref{U236e24}. It is a projection of the 4D PEL onto the $(c,a_4)$ plane, {\it i.e.}, each energy point in the $(c,a_4)$ map is minimized with respect to the non-axial $\eta$ and reflectional $a_3$ deformation parameters. The ground state (g.s.), first (A), and second (B) saddle points are marked in the plot. Beyond the second saddle B, two separate paths develop, an asymmetric one and a symmetric one. The exit points from the fission barrier leading to the asymmetric (C) and symmetric (D) fission valleys, are also marked. The upper value of the neck parameter $a_4=0.72$ corresponds to the neck radius approximately equal to the nucleon radius $r_{\rm neck}=r_0$ which we assume in the following as the scission criterion. The non-axial degree of freedom is important at a smaller elongation of the nucleus until the neighborhood of the second saddle. At larger deformation, its effect is negligible, allowing us to restrict the Langevin calculations to 3D when discussing fission dynamics. Moreover, the role of the higher-order deformation parameters $a_5$ and $a_6$ is rather small even in the region of well-separated fission fragments, as it was shown in Ref.~\cite{KDN21}. The ($c,\,{\rm A_h}$) cross-section of the PEL of $^{236}$U at $a_4=0.72$ is presented in Fig.~\ref{U236e23}. This cross-section corresponds roughly to scission ($r_{\rm neck}\simeq r_{\rm n}$), as noted above. Here ${\rm A_h}$ is the heavy fragment mass number. The close-to-scission configuration of the asymmetric valley evidenced in Fig.~\ref{U236e24} corresponds to the minimum at ${\rm A_h}=140$ and $c=2.2$, while the end of the symmetric valley of Fig.~\ref{U236e24} occurs at $c=2.83$. As expected, asymmetric fission of uranium leads to a more compact scission configuration as compared to symmetric splitting.
\begin{figure}[htb]
\includegraphics[width=0.5\textwidth]{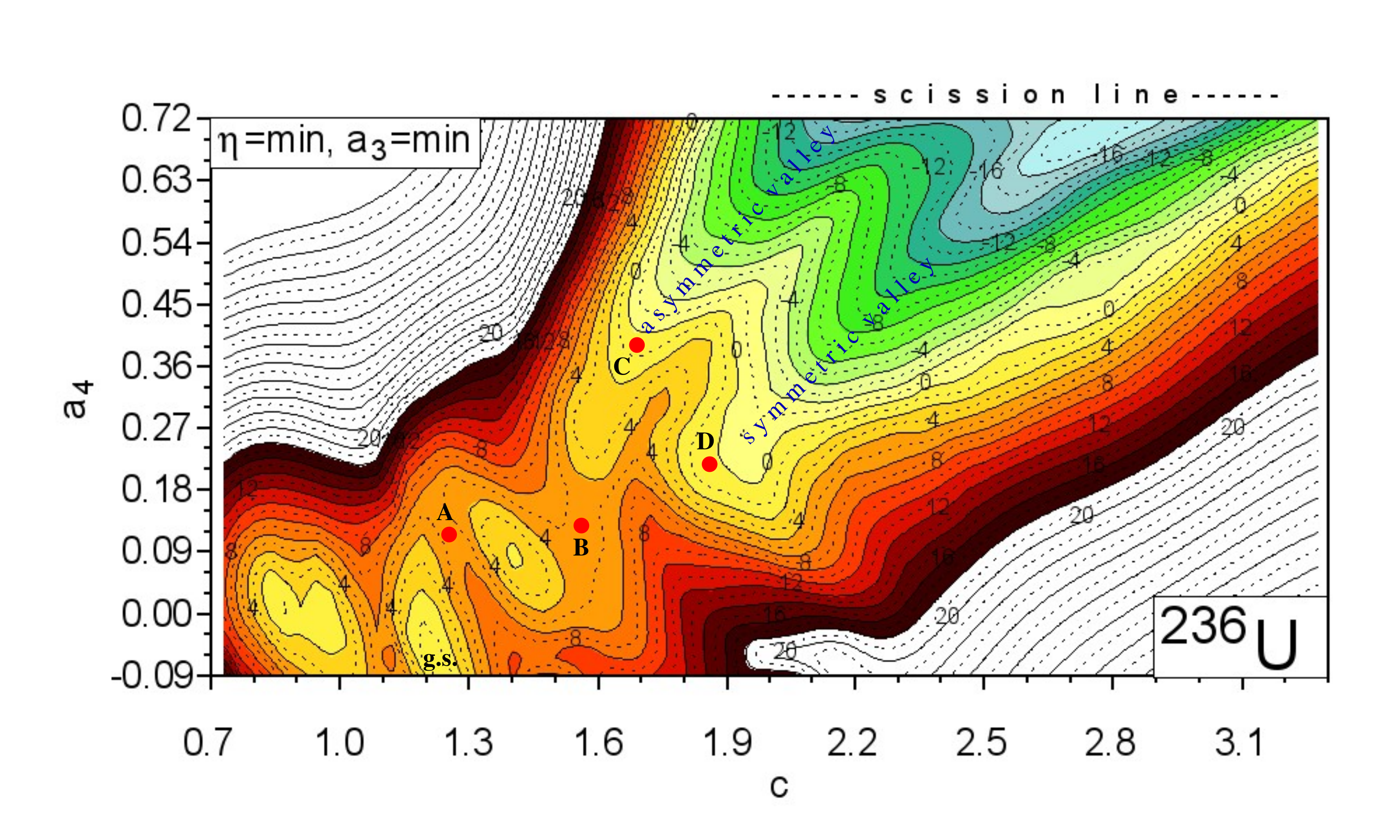}\\[-4ex]
\caption{Potential energy surface of $^{236}$U on the ($c,\,a_4$) plane. Each point is minimized with respect to the non-axial ($\eta$) and the reflectional ($a_3$) deformations.}
\label{U236e24}
\end{figure}
\begin{figure}[htb]
\includegraphics[width=0.5\textwidth]{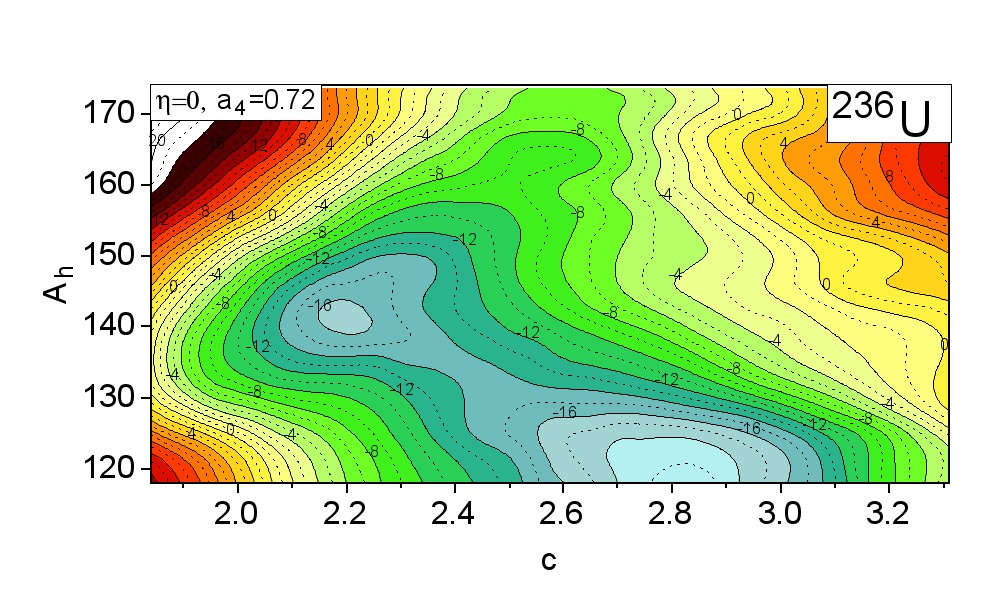}\\[-4ex]
\caption{Potential energy surface of $^{236}$U around the scission configuration ($a_4=0.72$) on the ($c,\,{\rm A_h}$) plane. Each point is minimized with respect to the non-axial ($\eta$) deformations.}
\label{U236e23}
\end{figure}


\subsection{Dynamical evolution}
\label{Sec2.1}

The Langevin equation governs the dissipative fission dynamics. In the generalized coordinates ($\{q_i\},~~i=1,2,...,n$) it has the following
form \cite{KPo12}:
\begin{equation}
\begin{array}{ll}
{dq_i\over dt} =& \sum\limits_{j} \; [{\cal M}^{-1}(\vec q\,)]_{i\, j} \; p_j  \\
 {dp_i\over dt}=& - {1\over 2} \sum\limits_{j,k} \,
           {\partial[{\cal M}^{-1}]_{jk}\over\partial q_i}\; p_j \; p_k
          -{\partial V(\vec q)\over\partial q_i}\\
          &- \sum\limits_{j,k} \gamma_{ij}(\vec q) \;
           [{\cal M}^{-1}]_{jk} \; p_k + F_i(t) \,\,, 
\end{array}
\label{LGV}
\end{equation}
Here $V(\vec q\,)=E_{\rm pot}(\vec q\,)-a(\vec q\,)T^2$ is the free-energy of fissioning nucleus having temperature $T$ and the single-particle level density $a(\vec q\,)$. The potential energy $E_{\rm pot}(\vec q\,)$ at a given deformation point ($\vec q$) is given by the macroscopic-microscopic prescription quoted in the previous section, and the level density $a(\vec q\,)$ at corresponding deformation is taken from Ref.~\cite{NPB02}. The  inertia ${\cal M}_{jk}$ and the friction $\gamma_{ij}$ tensors are evaluated in the irrotational flow and the wall approximation, respectively, as described in Refs.~\cite{BNP19,KDN21}.
\begin{figure}[htb]
\centerline{\includegraphics[width=0.4\textwidth]{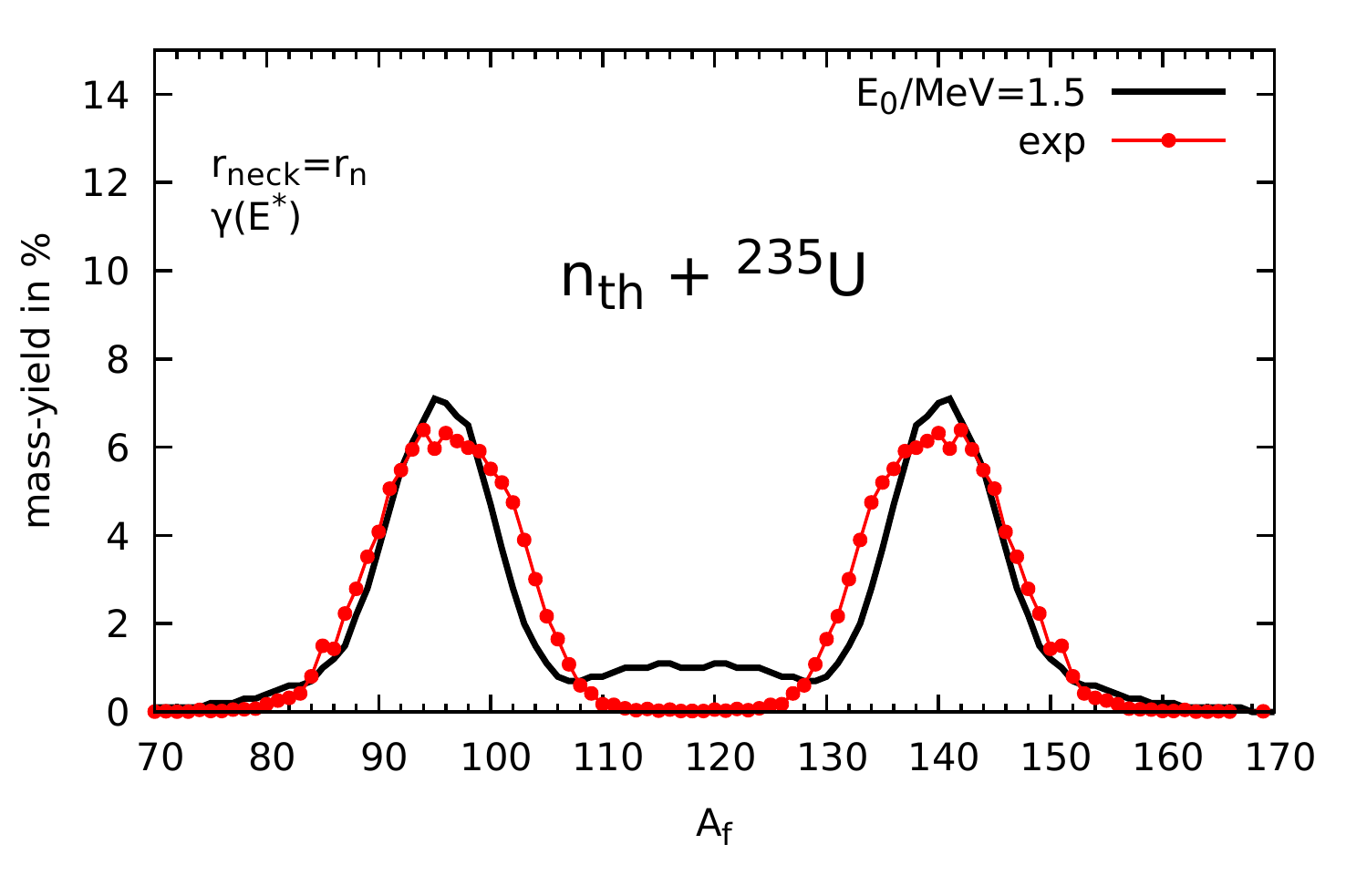}}
\caption{Fission fragment mass yield of ${\rm n_{th}\,+\,^{235}U}$ as a function of the mass of the fragment. The experimental data are taken from Ref. \cite{DKT69}.}
\label{U236th-my}
\end{figure}

The vector $\vec F(t)$ stands for the random Langevin force, which couples the collective dynamics to the intrinsic degrees of freedom and is defined as:
\begin{equation}
F_i(t) \!\!=\!\! \sum_{j} g_{ij}(\vec q\,) \; G_j(t) \,\,,
\label{rforce}
\end{equation}
where $\vec G(t)$ is a stochastic function whose strength $g(\vec q\,)$ is given by the diffusion tensor ${\cal D}(\vec q\,)$ defined by the generalized Einstein relation:
\begin{equation}
{\cal D}_{ij} \!\!=\!\!T^*\gamma_{ij} \!\!=\!\! \sum_{k} g_{ik} \; g_{jk}~,
\label{Eirel}
\end{equation}
where
\begin{equation}
T^*=E_0/{\rm tanh}\left({E_0\over T}\right)~.
\label{Tstar}
\end{equation}
Here $E_0= 3\times 0.5$ MeV is the zero-point collective energy. The temperature $T$ is obtained from the thermal excitation energy $E^*$ defined as the difference between the initial ($E_{\rm init}$ ) and the total collective energy, being the sum of the kinetic ($E_{\rm kin}$) and potential ($V$) energies of the fissioning nucleus at a given deformation point ($\vec q$)\\[-2ex] 
\begin{equation}
 a(\vec q\,)T^2=E^*(\vec q\,)=E_{\rm init}-(E_{\rm kin}+V)~.
\label{temp}
\end{equation}
For a given fissioning system, several thousands of Langevin trajectories leading to scission are run. From such samples, the properties of the primary fragments are evaluated, and at first place the mass and kinetic energy distributions presented below. 

\subsubsection{Mass yields}

The primary, or so-called pre-neutron, fission fragment mass yield as obtained for thermal neutron-induced fission of $^{235}$U is shown in Fig.~\ref{U236th-my}. Note that it was assumed here that each Langevin trajectory begins randomly at the region of the 2nd saddle (B) with the half-width of the initial distribution equal to the distance between the mesh-point ($\delta q_i$=0.03). It was observed that this leads to a predicted mass yield of $^{236}$U which is almost independent on the starting point: similar mass distributions are obtained when starting from the ground state deformation or from the first saddle (A). Our result describes pretty well the maxima and the tails of the experimental mass yield at large asymmetry \cite{DKT69}. However, the yield at symmetric is slightly overestimated.


\subsubsection{Total kinetic energy}

For each Langevin trajectory, the total kinetic energy (TKE) of the  fragments $E^{\rm frag}_{\rm kin}$ is given by the sum of the Coulomb repulsion energy ($V_{\rm Coul}$), the nuclear interaction energy of fragments ($V_{\rm nuc}$), and the pre-fission kinetic energy of the relative motion ($E^{\rm Coll}_{\rm kin}$) evaluated at the scission point ($q_{\rm sc}$):
\begin{equation}
E^{\rm frag}_{\rm kin}=V_{\rm Coul}(q_{\rm sc})
            + E^{\rm coll}_{\rm kin}(q_{\rm sc})+ V_{\rm nuc}(q_{\rm sc})~.
\label{Ekin}
\end{equation}  
The Coulomb repulsion energy is equal to the difference between the total Coulomb energy of the nucleus at the scission configuration and the Coulomb energies of both deformed fragments: 
\begin{equation}
V_{\rm Coul}={3e^2\over 5r_0}\left[{Z^2\over A^{1/3}}B_{\rm C}(\vec q_{\rm sc})
         -{Z_{\rm h}^2\over A_{\rm h}^{1/3}}\,B_{\rm C}(\vec q_{\rm h})
         -{Z_{\rm l}^2\over A_{\rm l}^{1/3}}\,B_{\rm C}(\vec q_{\rm l})\right]~,
\label{VCoul}
\end{equation}  
where $r_0=1.217\,$fm is the same charge radius as in the LSD mass-formula \cite{PDu09} and $B_{\rm C}$ is the ratio of the Coulomb energies of the deformed and spherical nucleus.

The nuclear interaction between the fragments at the scission point is approximately equal to the change of the nuclear surface energy when the neck breaks:
\begin{equation}
\begin{array}{rl}
V_{\rm nuc}(q_{\rm sc}) &= {\displaystyle  - 2\times E_{\rm surf}(0)
   \frac{\pi r^2_{\rm neck}({\rm sc})}{4\pi R_0^2}}\\[2ex]
   &= {\displaystyle -{1\over 2} E_{\rm surf}(0)
   \left(\frac{r_{\rm neck}}{R_0}\right)^2~.}
\end{array}
\label{Vnuc}
\end{equation}
Here $E_{\rm surf}= b_{\rm surf}A^{2/3}$, where $b_{\rm surf}$ is the surface tension LD coefficient. For $r_{\rm neck}=r_0$ and the nucleus radius $R_0=r_0A^{1/3}$ one obtains: $V_{\rm nuc}(q_{\rm sc})=-{1\over 2}b_{\rm surf}$, i.e., $V_{\rm nuc}(q_{\rm sc})\approx -9$ MeV for the neck-radius equal to the nucleon radius. We note that this prescription for  $E^{\rm frag}_{\rm kin}$ is undoubtedly a more accurate estimate of the fission-fragment kinetic energy than the frequently used point-charge approximation: $E_{\rm kin}=e^2 Z_{\rm h}Z_{\rm l}/R_{12}$, where $R_{12}$ is the distance between the fragment mass-centers.\\
The mean TKE as a function of fragment mass as obtained from the model is compared in Fig.~4 with the experimental data \cite{ATJ20}. These are reproduced well on the average. Some discrepancy is though to be noted. First, the predicted TKE around $A_h$=140 is too large. The yield in this mass region is nevertheless well described, see Fig.~\ref{U236th-my}. Thus, we ascribe the discrepancy in TKE as due to the limitation of the 4D parametrization to describe the scission shapes characteristic of the so-called Standard II mode, corresponding to a deformed heavy fragment and a slightly or even close to spherical light partner \cite{ANS18}. Second, the maximum of the calculated TKE, expected to occur for the Standard I mode with a heavy fragment in the vicinity of $^{132}$Sn, is seen to be shifted to larger masses around ${\rm A_h=136}$. The reason for this discrepancy is too fold: (i) the difficulty to describe in a 4D deformation space the compact shapes characteristic of Standard I mode, and (ii) the too large contribution of the symmetric mode, noted already in Fig.~\ref{U236th-my}, in the $A_h \approx$ 130 region, which corresponds to very elongated scission shapes, and thus lower the average TKE in this region.


\subsection{Charge equilibration at scission}

At the end of the Langevin trajectory, once the system has reached the scission point, the mass of the two fragments is determined by integrating the volume of the shapes at the left and right of the point of rupture, respectively. In the wide majority of macroscopic-microscopic models available on the market, the isotopic composition, equivalently N/Z ratio, of the fragments is next assumed to be identical to the one of the fissioning nucleus (see {\it e.g.} \cite{ACI14, RMo13, ACD21, LCW21, KDN21}). The UCD assumption was recently levelled off by M\"oller and Ichikawa \cite{MIc15} in a "6D" model by computing the probabilitiy of proton transfer between the two fragments along the dynamical evolution. In the fully microscopic approach, neutron and proton sharing at scission can in principle be obtained from the corresponding density distributions, see {\it e.g.} Ref.~\cite{verriere:2021} for a recent discussion. In the present work, we go beyond the UCD assumption which we employed in our previous model \cite{ LCW21, KDN21} as follows.\\
Starting from the fragment deformation at scission, we determine for each fragment mass the most probable charge based on the LSD energy and the pairing correlation energy. Such charge equilibration can be determined by looking at the change of the total energy of the fissioning system with the charge number of the heavy fragment $Z_{\rm h}$:
\begin{equation}
\begin{array}{rl} 
E(Z,A,Z_{\rm h}&;A_{\rm h},\vec q_{\rm h},\vec q_{\rm l})
    =E_{\rm LSD}(Z_{\rm h},A_{\rm h};\vec q_{\rm h})\\[+1ex]
    &+\,E_{\rm LSD}(Z-Z_{\rm h},A-A_{\rm h});\vec q_{\rm l})\\[+1ex]
    &+\,e^2Z_{\rm h}(Z-Z_{\rm h})/R_{12}-E_{\rm LD}(Z,A;0)~,
\end{array}
\label{echeq}
\end{equation}
where $Z, A$, and $Z_{\rm h}, A_{\rm h}$ are the charge and mass numbers of the mother nucleus and the heavy fragment, respectively. The mass as well as the deformation parameters of the heavy ($A_{\rm h},\,\vec q_{\rm h}$) and the light fragments ($A_{\rm l},\,\vec q_{\rm l}$) are given by the division of the volume according to the shape of the nucleus at scission at the end of the Langevin trajectory, as mentioned in the previous work \cite{ LCW21, KDN21}.
\\
The total energy as a function of the fragment charge number is shown in the upper panel of Fig.~\ref{qeqlsd}.

\begin{figure}[b!]
\begin{center}
\includegraphics[width=0.3\textwidth]{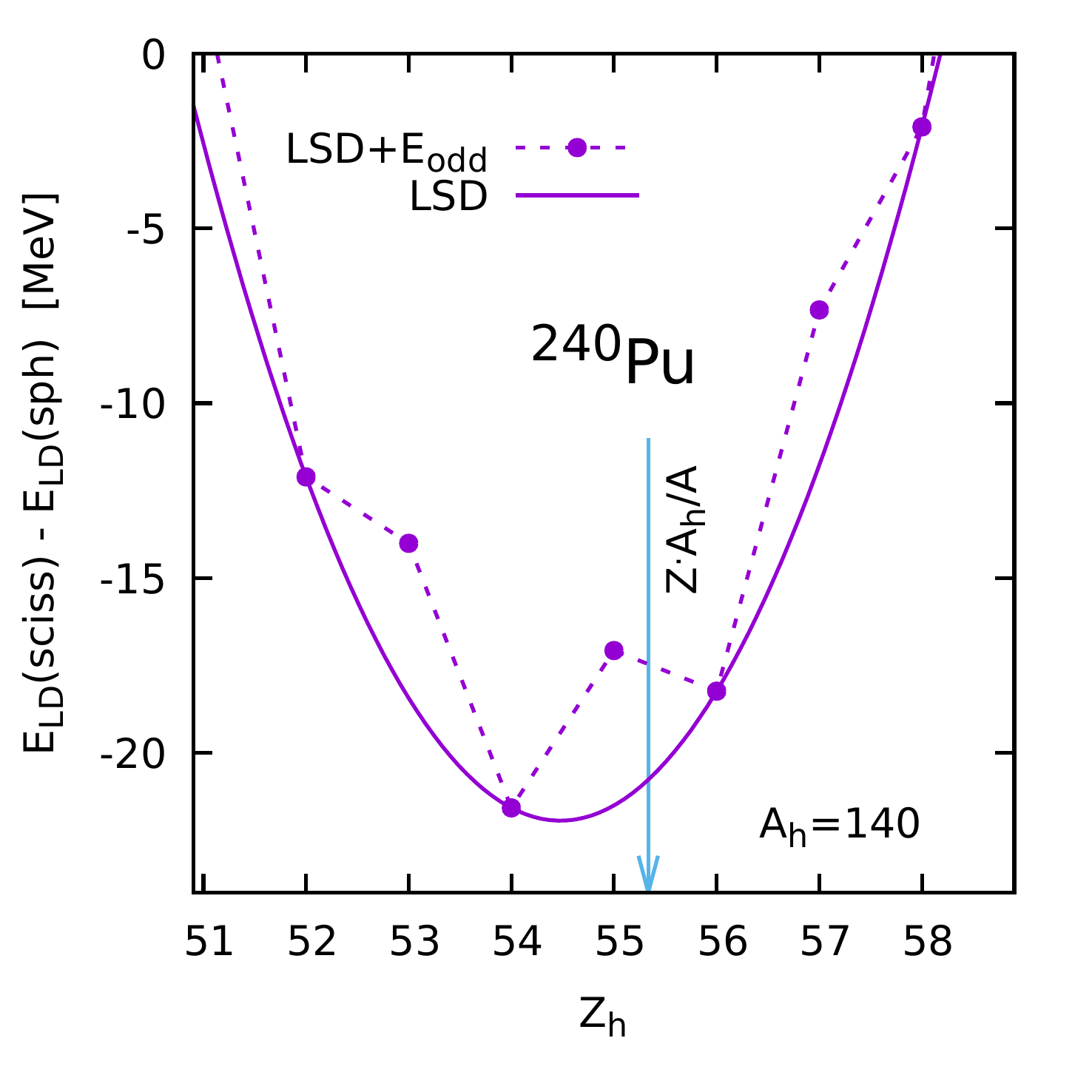}
\includegraphics[width=0.3\textwidth]{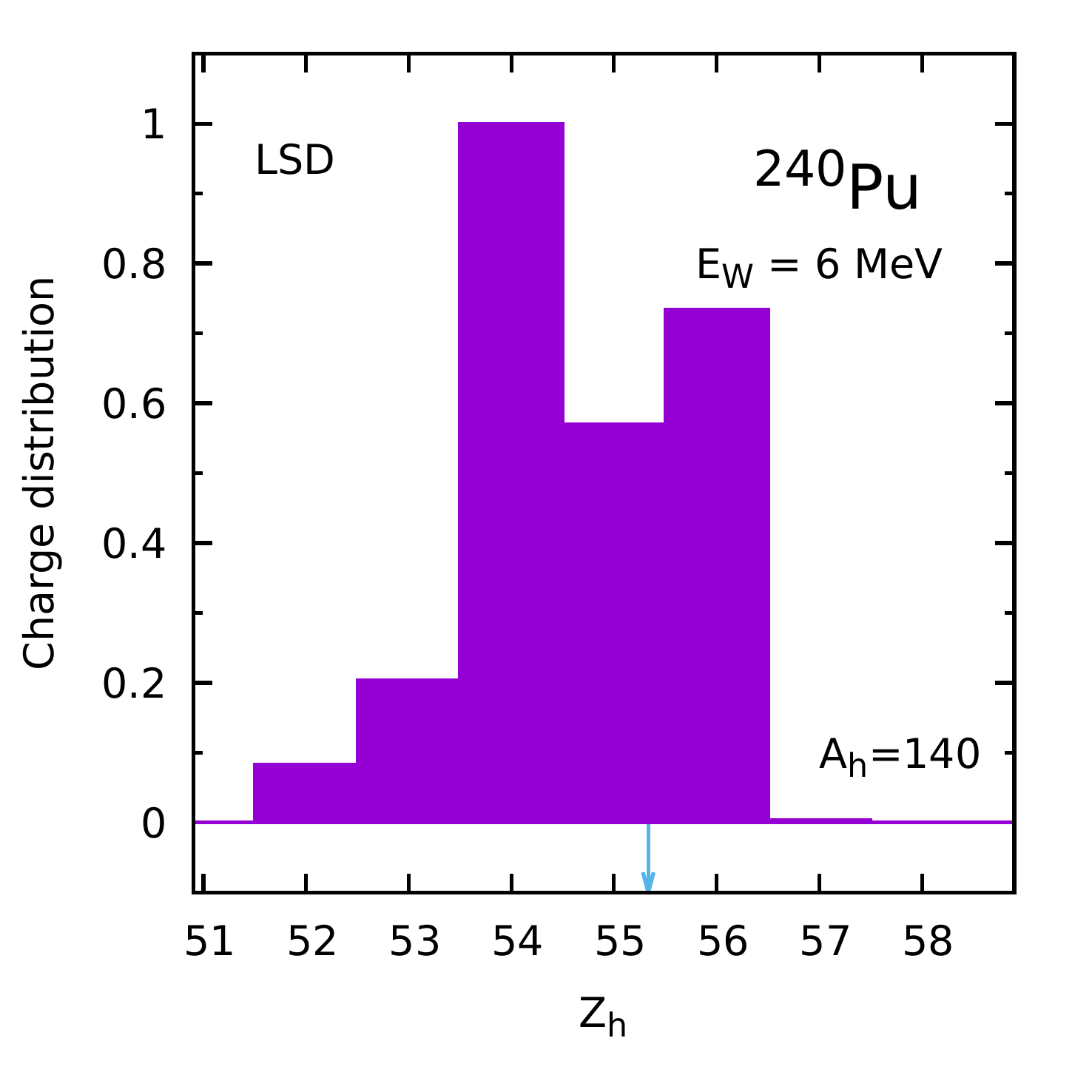}\\[-4ex]
\end{center}
\vspace{-1cm}
\caption{Energy of $^{240}$Pu at scission as a function of the heavy fragment charge number in the LSD mass formula \cite{PDu09} (top) and the Wigner distribution probability of the fragment charge number (bottom).}
\label{qeqlsd}
\end{figure}
The distribution of the heavy-fragment charge number can be estimated using a Wigner function corresponding to the energy $E$ given by Eq.~\ref{echeq} for different values of $Z_{\rm h}$:
\begin{equation}
 W(Z_{\rm h})=\exp\{-[E(Z_{\rm h})-E_{\rm min}]^2/E_{\rm W}^2]~,
\label{Wigner}
\end{equation}
which gives the distribution probability of the fragment charge shown in the bottom panel of Fig.~\ref{qeqlsd}. $E_{\rm min}$ in Eq.~\ref{Wigner} is the lowest discrete energy as a function of $Z_{\rm h}$. Furthermore, the following random number decides on the charge number $Z_{\rm h}$ of the heavy fragment, with $Z_{\rm l}=Z-Z_{\rm h}$. The energy $E_{\rm W}$ should be comparable with the energy distance $\hbar\omega_0$ between harmonic oscillator shells since we have a single-particle (proton) transfer, here, between the touching fragments.
\begin{figure}[t!]
\begin{center}
\includegraphics[width=0.4\textwidth]{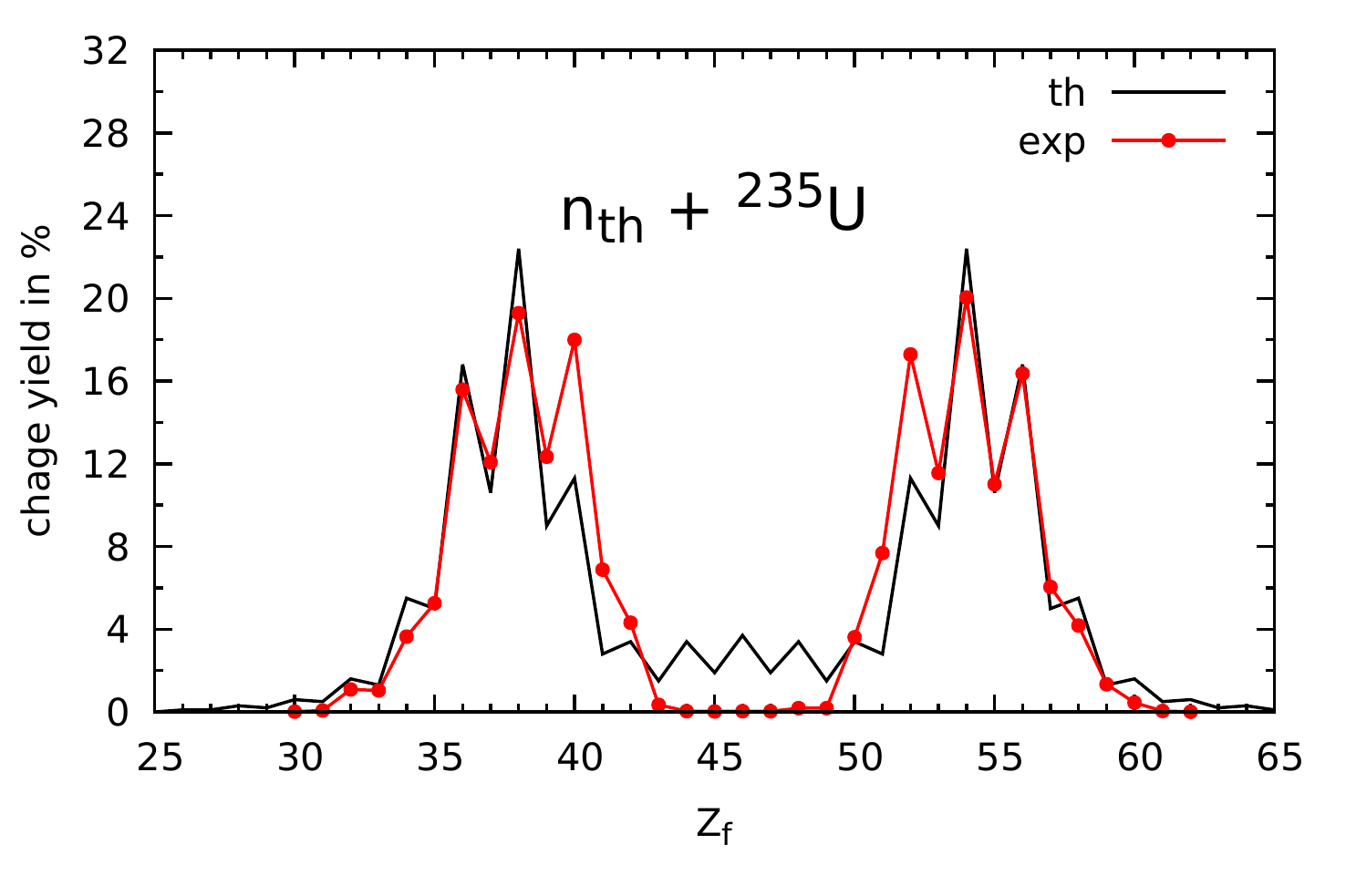}
\end{center}
\caption{Fission fragment charge yield of ${\rm n_{th}\,+\,^{235}U}$. The experimental data (red points) are taken from Ref. \cite{MSc17}.}
\label{U236th-chy}
\end{figure}

The above outlined prescription permits to go beyond the UCD hypothesis by accounting for charge equilibration for a given mass split.  The resulting fission fragment charge yield is compared with the data \cite{MSc17} in Fig.~\ref{U236th-chy}. As one can see the odd-even effect for the most probable fission fragment elements are well reproduced with  our simple model which is solely based on the LSD macroscopic energy, for the largest yields. Theory overestimates the staggering for most asymmetric splits and at symmetry. This dependence of the magnitude of the staggering with fragment charge is under vivid debate \cite{caamano:2011} due to its connection with the influence of shell effects and dissipation in fission \cite{ramos:2023}. Within the present modeling, it will be the subject of future development. We note that a similar procedure could be introduced to account for neutron pairing. However, since evaporation after scission widely washes it out, it is hardly seen in experiment, and not much exploitable.


\subsection{Post-scission evaporation}

The primary fragments produced right at scission are in general excited. They return to their respective ground state by emitting neutrons and $\gamma$-rays. Our previous model \cite{LCW21, KDN21} was extended to account for post-scission evaporation of neutrons. Competition with $\gamma$-ray emission has a negligible impact on neutron evaporation, as it occurs mostly below the fragment neutron separation energy. Inclusion of $\gamma$-ray emission is thus left for future development.\\
The excitation of the fissioning nucleus available at scission, and to be shared between the primary fragments, is evaluated as specified above with Eq.~\ref{temp}. It is then assumed  that the thermal energy of a given fragment $E^*_i$ at the  scission point is proportional to its single-particle level density:
\begin{equation}
 {E^*_{\rm l}\over E^*_{\rm h}}= {a(Z_{\rm l},A_{\rm l};{\rm def}_{\rm l})\over a(Z_{\rm h},A_{\rm h};{\rm def}_{\rm h})}
\label{Eteq}
\end{equation}
 with $E^*=a(\vec q)\,T^2=E^*_{\rm l}+E^*_{\rm h}$ is given by Eq.~\ref{temp}.

Since the fragments have usually a deformation at scission which differs from their equilibrium configuration, they very fast relax to the ground-state shape. The deformation energy released by this relaxation is transformed into excitation energy.  The deformation energy of each fragment can be evaluated in the LD model \cite{PDu09}:
\begin{equation}
E_{\rm def}^{(i)}\approx E_{\rm LD}(Z_i,A_i,{\rm def}_i)
                 - E_{\rm exp}(Z_i,A_i,{\rm g.s.})~.
\label{Edefi}
\end{equation}
The total excitation energy ($E^{(i)}_{\rm exc}$) of fragment $i$ is then the sum of its thermal and deformation energies:
\begin{equation}
 E_{\rm exc}^{(i)}=E_{\rm def}^{(i)}+E_i^*=a(i)\,T_i^2~.
\label{Eexci}
\end{equation}
For each fragment, this excitation energy is available for neutron emission.

The maximal energy of a neutron emitted from a fragment (mother) can be obtained from the energy conservation law:
\begin{equation}
\epsilon_{\rm n}^{\rm max}= M_{\rm M}+E_{\rm M}^*-M_{\rm D}-M_{\rm n}~,
\label{En}
\end{equation}
where $M_{\rm M},\,M_{\rm D},\,M_{\rm n}$ are the mass excesses of mother and daughter nuclei and of the neutron, respectively. These data can be taken from a mass table \cite{KWH21}. The thermal excitation energy of the daughter nucleus is: 
\begin{equation}
 E_{\rm D}^*=\epsilon_{\rm n}^{\rm max}-\epsilon_{\rm n}~.
\label{Eexc}
\end{equation}
Here $e_{\rm n}$ is the kinetic energy of the emitted neutron.
\begin{figure}[htb]
\begin{center}
\includegraphics[width=0.4\textwidth]{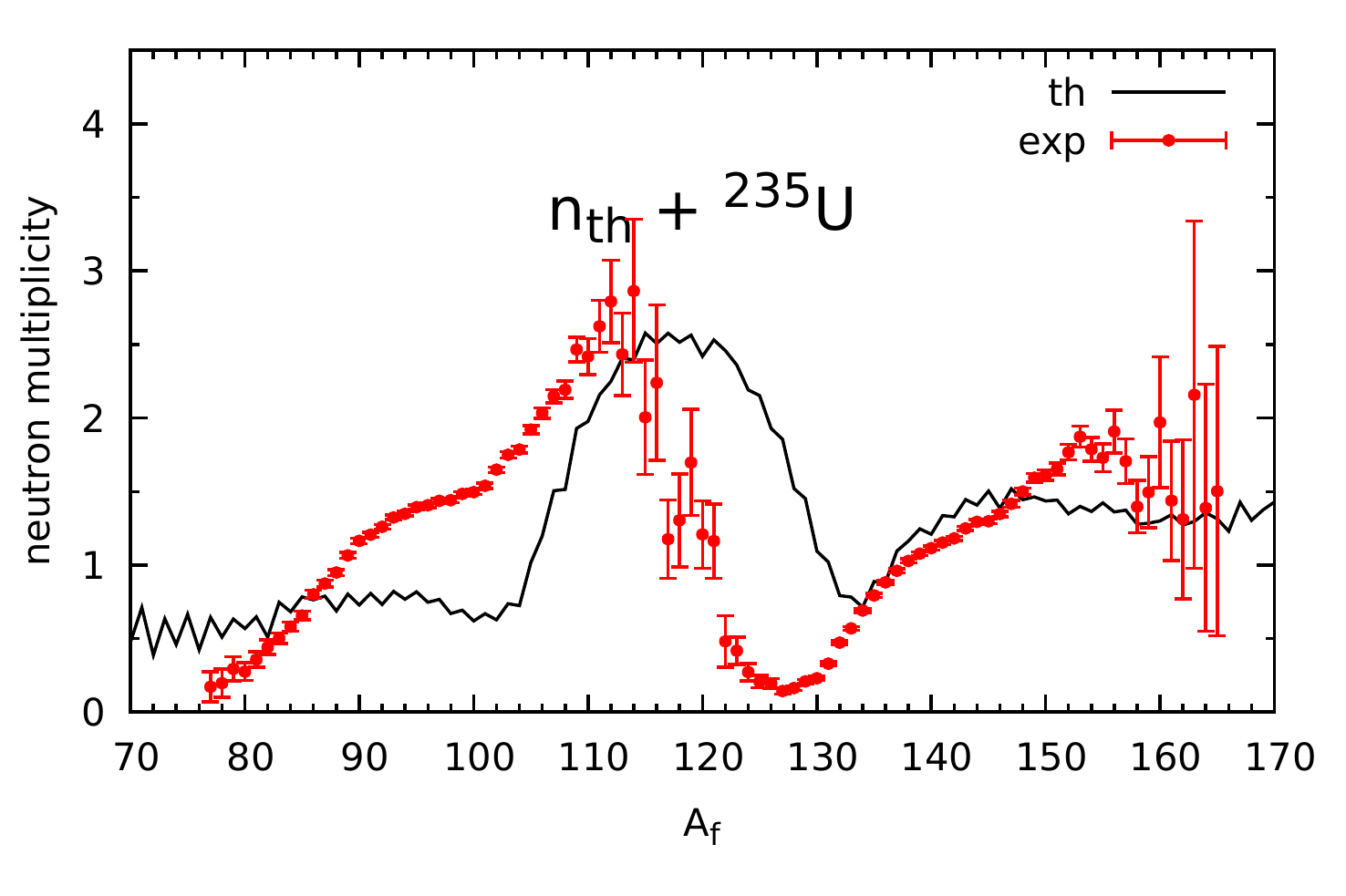}
\end{center}
\vspace{-1cm}
\caption{Post-scission neutron multiplicity as a function fragment mass for ${\rm n_{th}\,+\,^{235}U}$. The experimental data (red points) are taken from Ref.~\cite{GHO18}.}
\label{u236th-nm}
\end{figure}

The neutron emission probability for a (mother) fragment with excitation energy $E^*_{\rm M}$ is given by the Wei{\ss}kopf formula \cite{Del86}:
\begin{equation}
\Gamma_{\rm n}(\epsilon_{\rm n})={2\mu\over\pi^2\hbar^2\rho_{\rm M}
(E^*_{\rm M})}\int\limits_0^{\epsilon_{\rm n}}\sigma_{\rm inv}(\epsilon)\,
\epsilon\,\rho_{\rm D}(E^*_{\rm D})\,d\epsilon~.
\label{Wei}
\end{equation}
Here $\mu$ is the reduced mass of the neutron, $\sigma_{\rm inv}$ is the neutron inverse cross-section \cite{DFF80}:
\begin{equation}
\begin{array}{ll}
\sigma_{\rm inv}(\epsilon)&=[0.76+1.93/A^{1/3}\\ 
&+ (1.66/A^{2/3}-0.050)/\epsilon]\,\pi\,(1.7A^{1/3})^2~,
\end{array}
\label{ncross}
\end{equation}
while  $\rho_{\rm M}$ and $\rho_{\rm D}$ are, respectively, the level densities of mother and daughter nuclei:
\begin{equation}
\rho(E)={\sqrt{\pi}\over 12a^{1/4}E^{5/4}}\exp(2\sqrt{aE})~,
\label{lden}
\end{equation}
Like in other parts of the model, the single-particle level density parameters $a$ of the mother and the daughter are taken from Ref.~\cite{NPB02}.\\
\begin{figure}[b!]
\centerline{\includegraphics[width=0.4\textwidth]{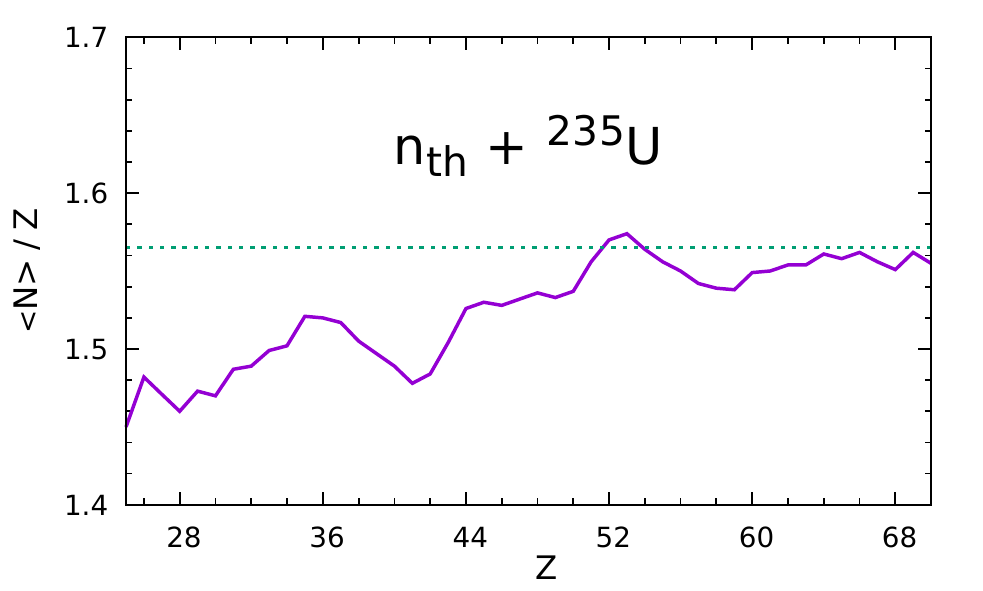}}
\caption{Average post-scission neutron to proton ${\rm <N>/Z}$ ratio for n$_{th}$+$^{236}{\rm U}$. The dashed line represents the ratio of the compound nucleus.}
\label{avnoz}
\end{figure}
Neutron evaporation is assumed to take place until the fragment reaches an excitation energy comparable to the neutron separation energy, for which we take an average value of 6 MeV (this energy is further exhausted by $\gamma$-rays as also observed in experiment, see {\it e.g.} Ref.~\cite{oberstedt:2014}).\\
The number of neutrons emitted by the fragments as function of their mass is displayed in Fig.~\ref{u236th-nm} and compared with the measurements \cite{GHO18}. The {\it sawtooth} shape observed in the experimental data is only roughly reproduced by the theoretical results. 
The too large multiplicity predicted in the range between A $\approx$ 116 and 130 is partly due to the too large amount of very elongated scission shapes originating from the LD fission mode in this region, as already discussed in Fig.~\ref{U236th-my}. The fragments of this mode experience a substantial shape relaxation after scission, what increases the excitation energy available for evaporation (see also Figs.~\ref{U236th-Ynz} and \ref{U236th-Esnz}). Furthermore, the too large amount of evaporation in the vicinity of $^{132}$Sn is additionally due to the limitation of the model to describe the specific shapes of the Standard I mode. The little under-prediction at A $\approx$ 155  may similarly point the issue of shape parametrization for those elongated heavy fragments. When the influence of structural effects in the heavy fragment dominates, energy minimization will naturally favor those pre-scission configurations which reproduces at best the shape of the heavy "side" of the mono-nucleus approaching scission. The limited number of collective coordinates, will necessarily bias the shape of the light counterpart and thus its excitation energy and post-scission evaporation. That partly explains the discrepancy between theory and experiment in the region A $\approx$ (90-110). It is expected that inclusion of higher deformation parameters, namely $a_5$ and $a_6$ which allow to better control the fragments deformation, will substantially improve the description of post-scission neutron multiplicities. 

The average fragment neutron to proton ${\rm <N>/Z}$ ratio after post-scission evaporation for fission of $^{236}{\rm U}$ at thermal energies is shown in Fig.~\ref{avnoz} as a function of the fragment charge number. The N/Z ratio of the initial system is given by the dashed line as for reference. The change of the fragment ${\rm <N>/Z}$ with respect to the mother nucleus is due to charge equilibration at scission and post-scission neutron evaporation}. The general behavior observed in experiment, see {\it e.g.} Ref.~\cite{RCF19}, with the heavy fragment being relatively neutron-rich and the lighter neutron-poor for fission of typical actinides, is reproduced. Though, the influence of shell effects in the vicinity of $^{132}$Sn is weaker in theory as compared to the measurement. As discussed above, we mostly attribute this to the limitation in the description of the particularly compact scission shapes characteristic of those fragmentations. The excitation of the heavy partner is then slightly overestimated, the neutron multiplicity gets too large, what lowers the ${\rm <N>/Z}$ ratio. To be best of our knowledge, apart from the present work, there are only two dynamical models which addressed the experimentally observed evolution of ${\rm <N>/Z}$ with fragment charge (or mass): While the enhanced "6D" macroscopic-microscopic model of M\"oller and Ichikawa \cite{MIc15} achieved a very good quantitative description \cite{schmitt:2021}, the description by the self-consistent model of Verriere et al. \cite{verriere:2021} remained qualitative only.\\
\\
The model developed in the present work calculates all (except the angular momentum) fragment properties in a consistent manner, and takes properly care of the correlations between the various quantities. For instance, the primary fragment N and Z distributions and associated shapes predicted by the calculation of the dynamical evolution up to the scission point determine the TKE. The primary (N, Z) population together with TKE gives the total excitation energy (TXE). The fragment deformation at scission together with the TXE enters the calculation of the intrinsic excitation energy of the fragments, which finally determine the neutron multiplicity and N/Z neutron excess. Correlations are essential to get further insight into the process, as well as to understand possible deviation between experiment and theory.   
  
\begin{figure}[b!]
\includegraphics[width=0.5\textwidth]{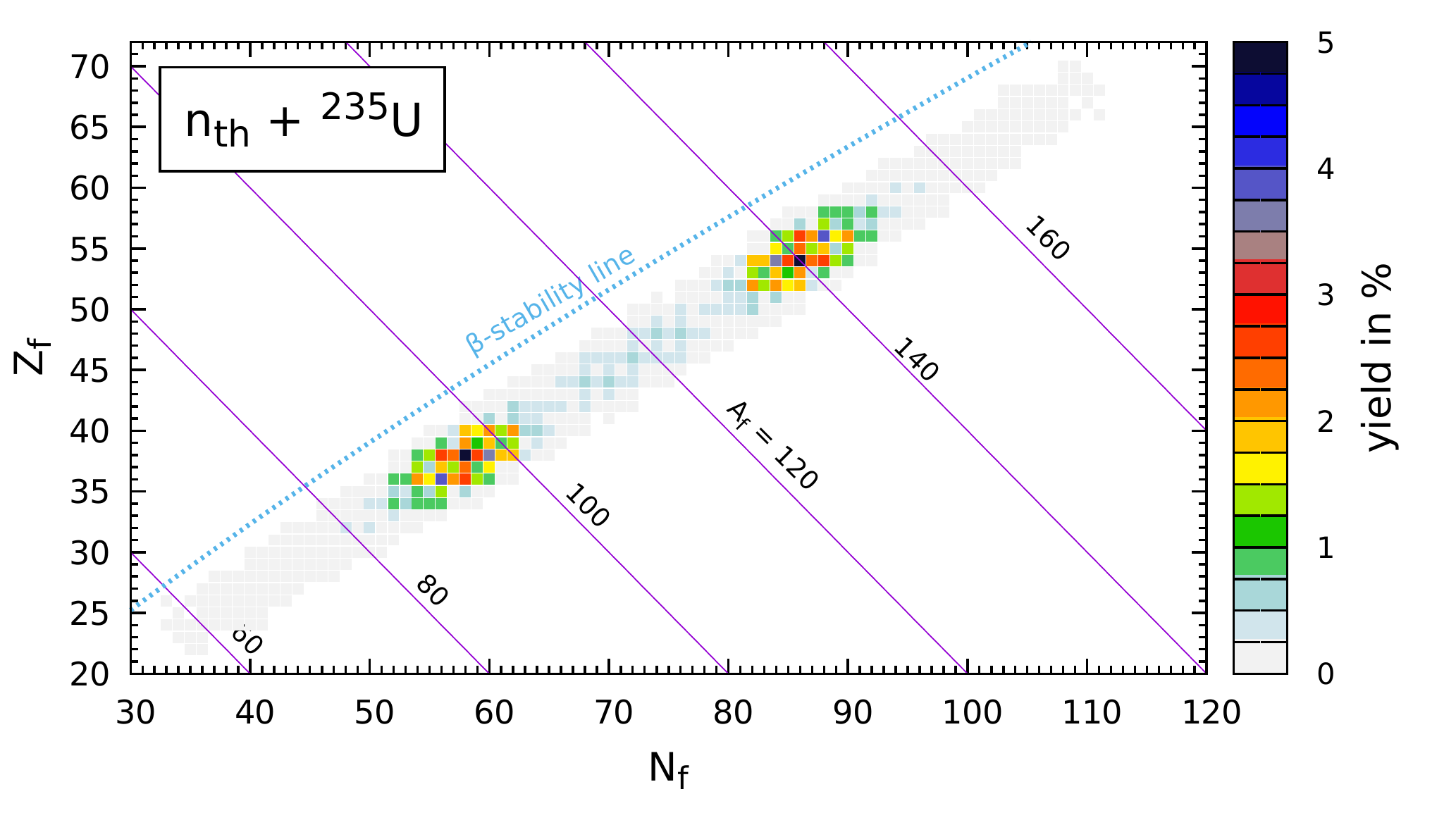}\\
\includegraphics[width=0.5\textwidth]{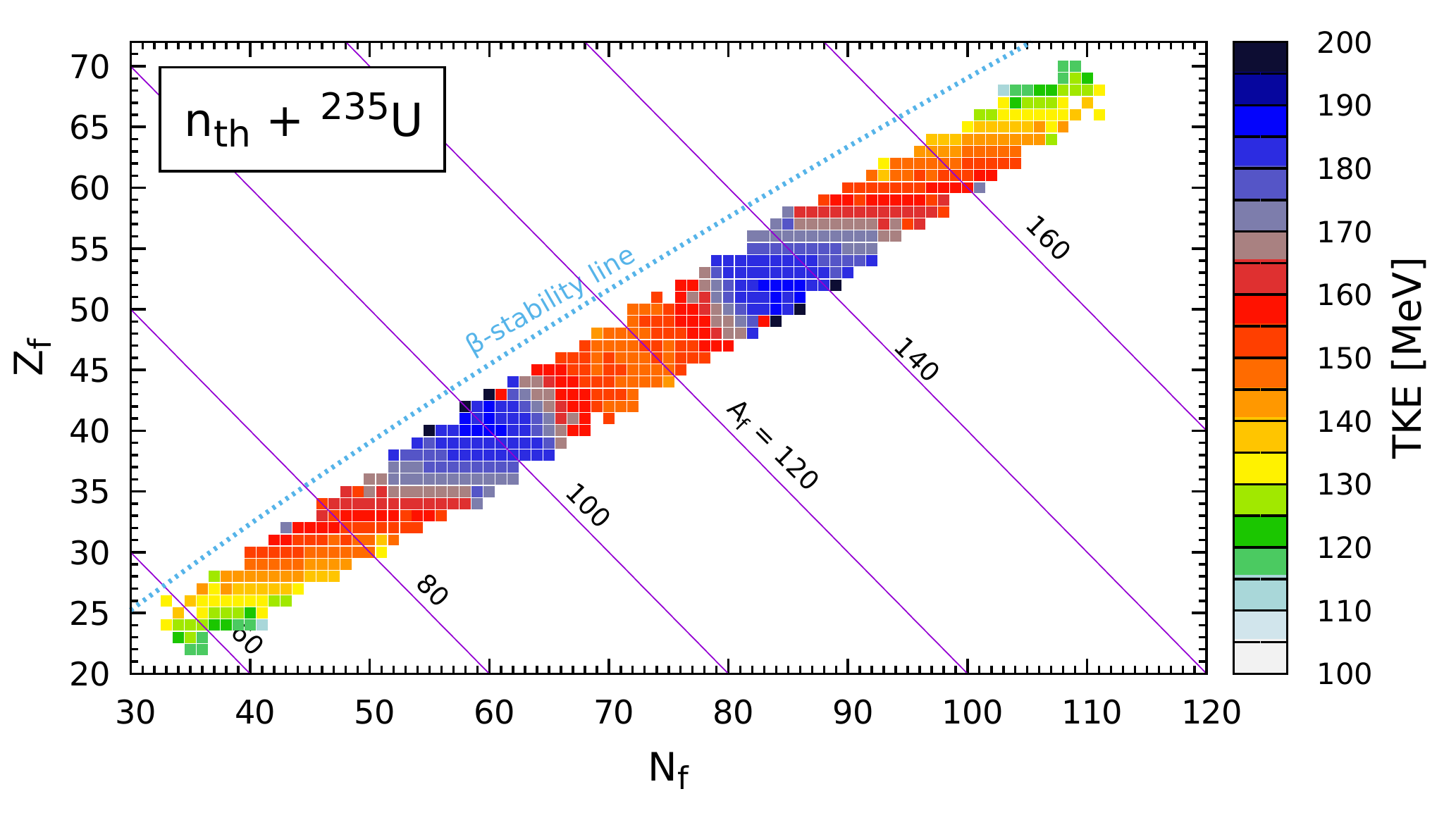}\\[-4ex]
\caption{Fission fragment yield (top) and TKE (bottom) for n$_{th}$+$^{235}$U on the $({\rm N_f,\,Z_f})$ plane.}
\label{U236th-Ynz}
\end{figure}
The primary fragment yield and TKE of $^{236}{\rm U}$  are shown in Fig.~\ref{U236th-Ynz} on the $({\rm N_f,\,Z_f})$ plane. In our model, the most probable primary  fragments are $^{140}$Xe and $^{96}$Sr, consistent with what suggested by combining the experimental observations of Refs.~\cite{DKT69, GPo86, bockstiegel:2008}. The largest TKE$\gtrsim$190 MeV corresponds to neutron-rich fragments with mass A between $\approx$130-140 and correlated with light fragments around A=100 having smaller neutron excess. Rather small values of the TKE of the fragments, equal to approximately 140 MeV, are calculated for symmetric fission. The larger TKE for the Standard I and II modes as compared to the LD symmetric mode well established from experiment \cite{bockstiegel:2008} is thus reproduced. Though, the measured difference between Standard I and Standard II is not evident in the calculation, presumably due to the limited number of collective coordinates. That translates into a fragment excitation in the vicinity of $^{132}$Sn which is somehow too large, and consequently an overestimation of the number of neutrons emitted, as noted above. This is indeed seen in Fig.~\ref{U236th-Esnz} which displays the fragment excitation energy and neutron multiplicity on the $({\rm N_f,\,Z_f})$ plane.
\begin{figure}[t!]
\includegraphics[width=0.5\textwidth]{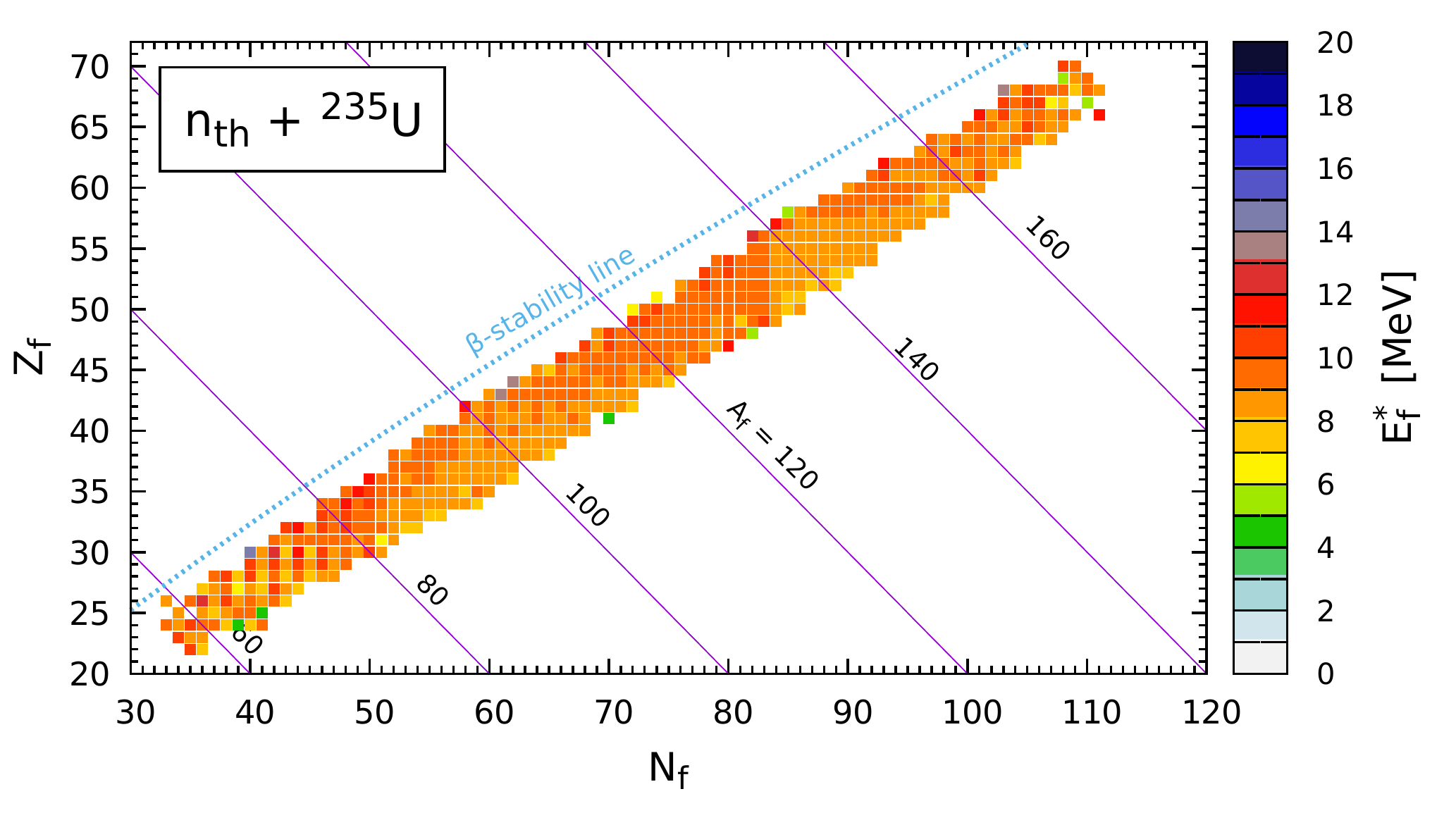}\\
\includegraphics[width=0.5\textwidth]{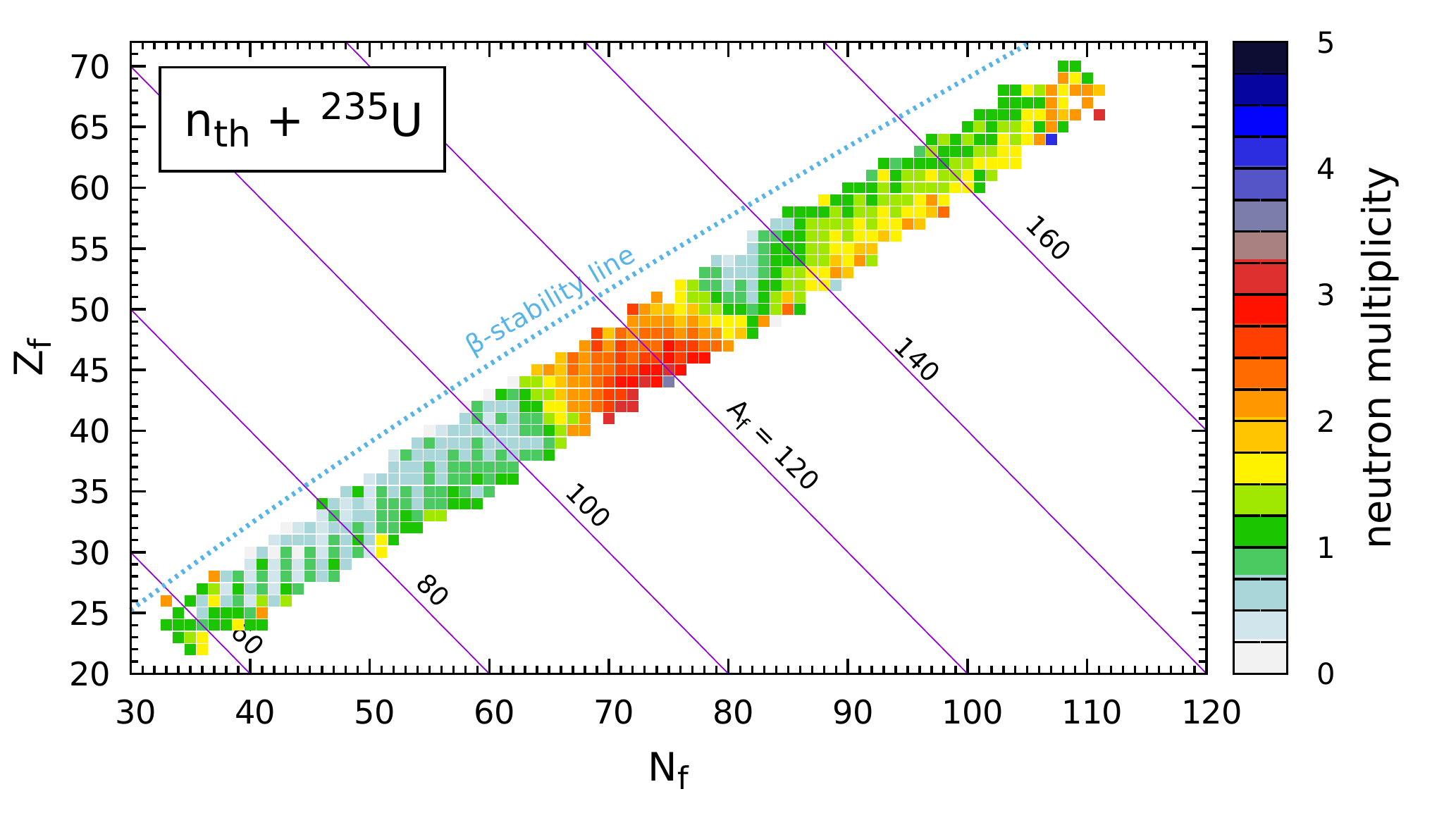}\\[-4ex]
\caption{Fission fragment excitation energy (top) and neutron multiplicity (bottom) for n$_{th}$+$^{235}$U  on the $({\rm N_f,\,Z_f})$ plane.}
\label{U236th-Esnz}
\end{figure}
\begin{figure}[t!]
\includegraphics[width=0.5\textwidth]{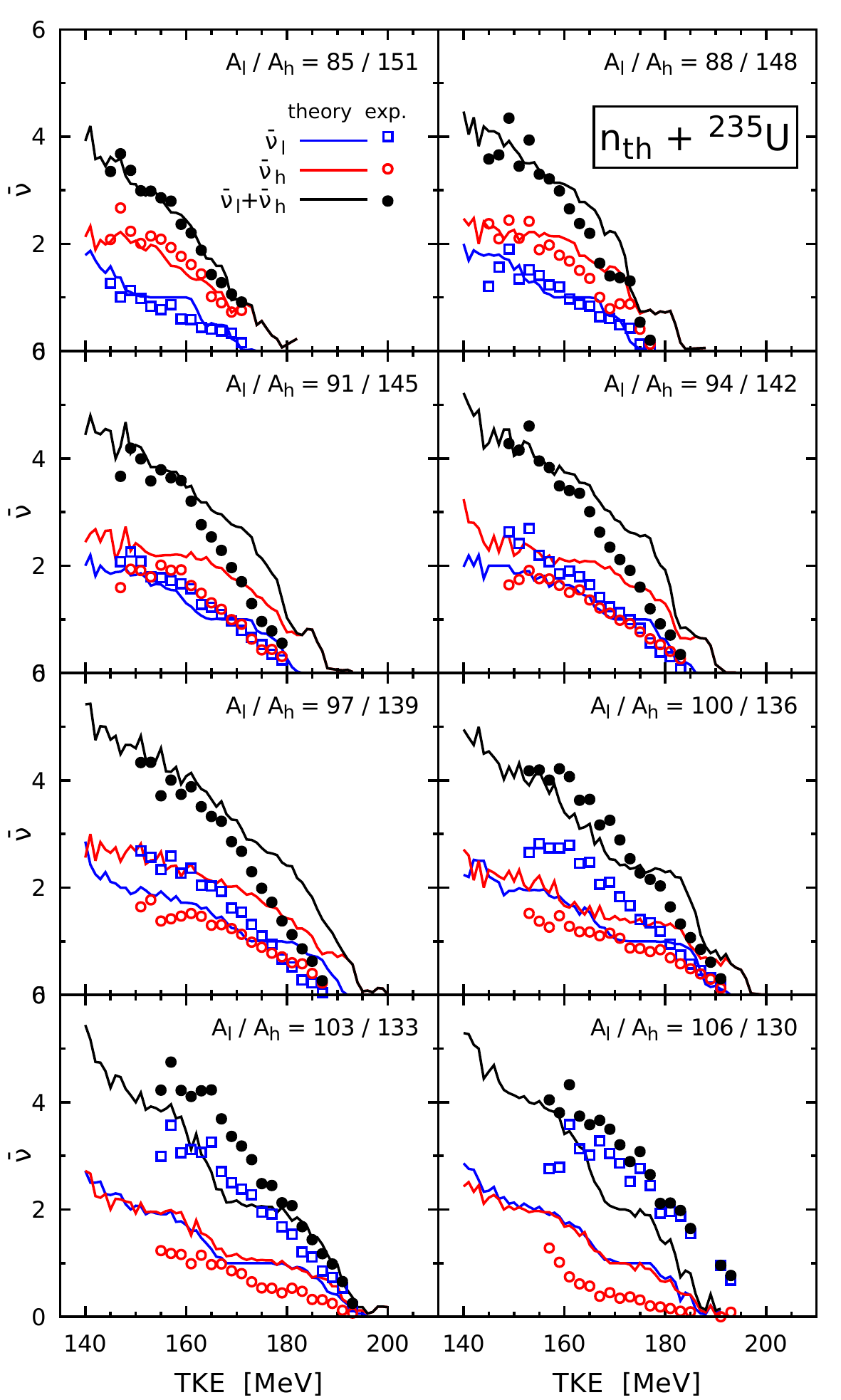}
\caption{Average neutron multiplicity as a function of TKE for selected mass pairs, as indicated in the top right corner of each panel. Experimental data of Ref.\cite{GHO18} for the light and the heavy fragment separately, and their sum, are compared to the calculation}.
\label{nmTKE}
\end{figure}
\\
\\
A further stringent test of the model is presented in Fig.~\ref{nmTKE}, where the experimental data on the average neutron multiplicity as a function of the fission fragment TKE for ${\rm n_{th}\,+\,^{235}U}$ \cite{GHO18} are displayed for various mass gates, and compared to the predictions of our model. The description is pretty good, except for those pairs of fragments substantially contributed by the Standard I mode. For the latter, the theoretical neutron multiplicity is too large for the heavy fragment, what is in line with the interpretation of the discrepancies observed above. Though, it is to be noted that, at the same time, the neutron multiplicity of the light partner is underestimated. That is mostly attributed to the impact of the aforementioned bias introduced by the restriction to 4 dimensions. Within the 5D Brownian shape motion model, Albertsson et al. \cite{ACD21} obtained a better description for these fragment pairs. That supports our conjecture that an increase in dimensionality of our model, with the inclusion of independent deformation variables for the light and heavy fragments ($a_5$ and $a_6$), will cure most of the deviation of the current theory. This conjecture is supported also by the analysis of the N/Z ratio reported above, where the "6D" model of Ref.~\cite{MIc15} based on the same 5D deformation landscape as Ref.~\cite{ACD21} achieves a better description than the present 4D model. Nevertheless, it is not excluded that part of the discrepancy observed here may be due to the prescription of excitation energy sharing and charge equilibration at scission. For both aspects we consider for the fragments the macroscopic energy, only (i.e. shell effects are omitted). Furthermore, unlike Ref.~\cite{ACD21}, we use an approximate formula \cite{RCF19} for the density of states of the deformed fragments, rather than the actual s.p. level densities with the shell effects.\\
The present investigation demonstrates that high-fold correlation data, which nowadays are becoming available in experiment, and when they are properly propagated in the calculation along the real-time evolution of the fissioning system, are crucial in order to evidence in an un-ambiguous manner the origin of possible weak points of a model. That is important to guide further development of the theory.


\section{Application to the Fm chain}
\label{sec_fm}

Experiment well established that the Fm isotopic chain exhibits a very peculiar trend in fragment properties with the size of the fissioning system: the fragment mass distribution changes abruptly from asymmetric for $^{256}$Fm to narrow and symmetric in $^{258}$Fm \cite{Hul86,Hul89}. At the same time, the TKE has a double-humped shape for the heavier isotope. This is certainly the best example of bimodal fission.  The first theoretical papers providing an explanation for the origin of this observation  appeared at the end of the 80s; they were all based on a static analysis of the PEL (see {\it e.g.} \cite{CRS89}). Thanks to the development of theory and increase in computing resources since then, advanced dynamical calculations are now possible within both the macroscopic-microscopic approach (see Refs.~\cite{miyamoto:2019, UII19, ACD21} for 3D, 4D and 5D models, respectively) and the microscopic self-consistent framework \cite{regnier:2019}. There is a wide consensus that the sudden transition observed along the isotopic chain of Fm (and of a few more trans-fermium elements) is caused by the proximity of strong shell effects at symmetry which fragments approach $^{132}$Sn with increasing fissioning isotope mass. As obvious from the quoted theoretical papers, a proper description of the mass and TKE yields along the Fm isotopic chain is a good test for any theoretical model. 
\begin{figure*}[t!]
\includegraphics[width=\textwidth]{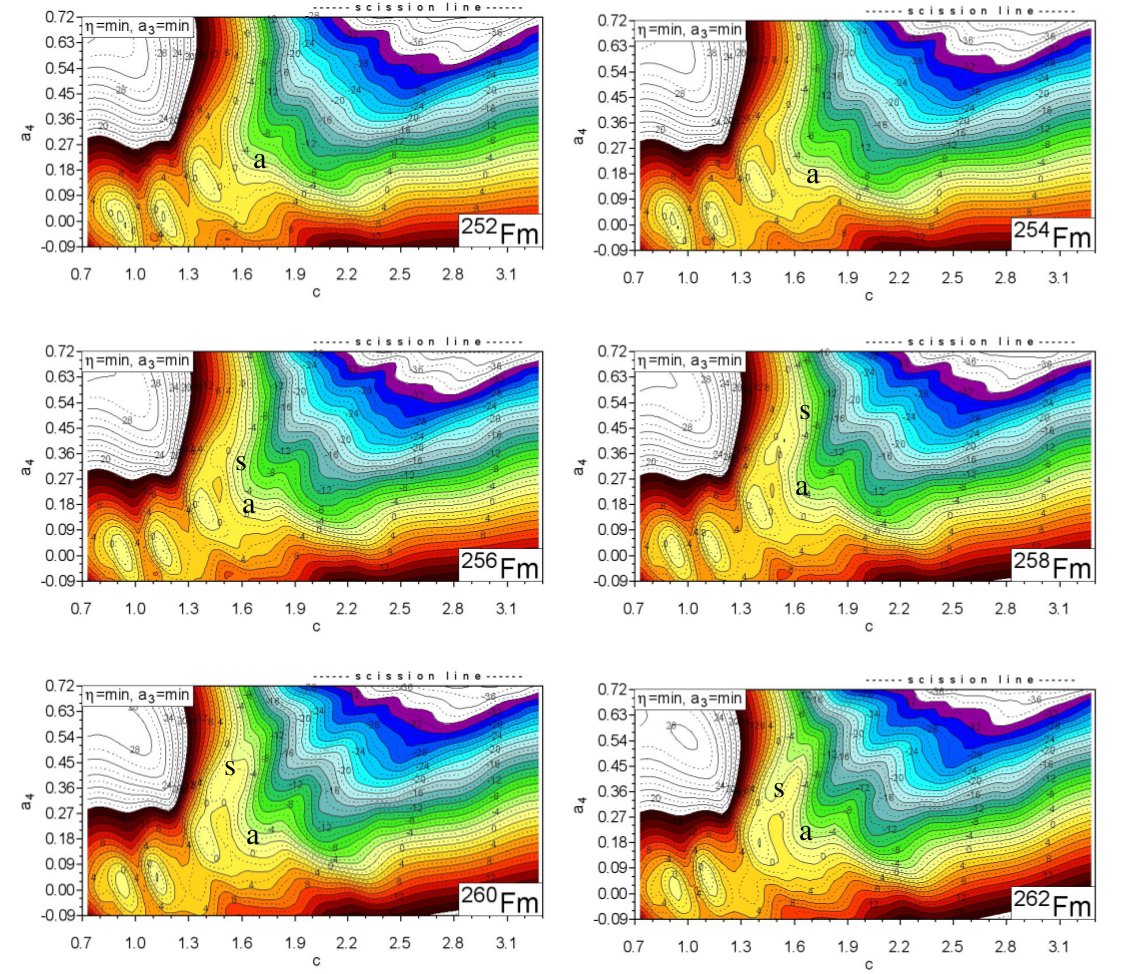}
\caption{Potential energy surface of the even-even $^{252-262}$Fm isotopes on the ($c,\,a_4$) plane.  Each point is minimized with respect to the non-axial ($\eta$) and the reflectional ($a_3$) deformation. The asymmetric {\bf a} and symmetric {\bf s}  exit points from the fission barrier are marked. }
\label{Fm-maps}
\end{figure*}

The model described in the present work was used to calculate the fission fragment properties (mass, charge, TKE, post-scission neutron multiplicity) along the Fm chain. All parameters were set identical to those employed in the previous section for thermal neutron-induced fission of $^{236}$U. The 4D PEL's of  the even-even $^{252-262}$Fm  isotopes projected onto the ($c,\,a_4$) plane are shown in Fig.~\ref{Fm-maps}. Each point of the maps is minimized with respect to the non-axial ($\eta$) and the pear-like ($a_3$) deformation as for $^{236}$U. For the lightest isotopes $^{252,254}$Fm, the outer saddle point is rather well defined and located at $c \approx$ 1.5 and $a_4 \approx$ 0.18. Its exit point, denoted {\bf a} in all maps, marks the beginning of a valley which corresponds to asymmetric fission (as seen from the corresponding minimized $a_3$; not shown here). Between $^{256}$Fm and $^{258}$Fm the pattern in the outer saddle region clearly changes, and still another outer saddle (at $c \approx$ 1.45 and $a_4 \approx$ 0.27) appears. A new fission valley develops beyond this additional outer barrier in $^{258}$Fm. It corresponds to compact symmetric fission configurations and is denoted {\bf s}. The PEL's of Fig.~\ref{Fm-maps} suggest that the symmetric valley might attract most of the flux for the heaviest Fm isotopes.
\begin{figure}[t!]
\begin{center}
\includegraphics[width=0.4\textwidth]{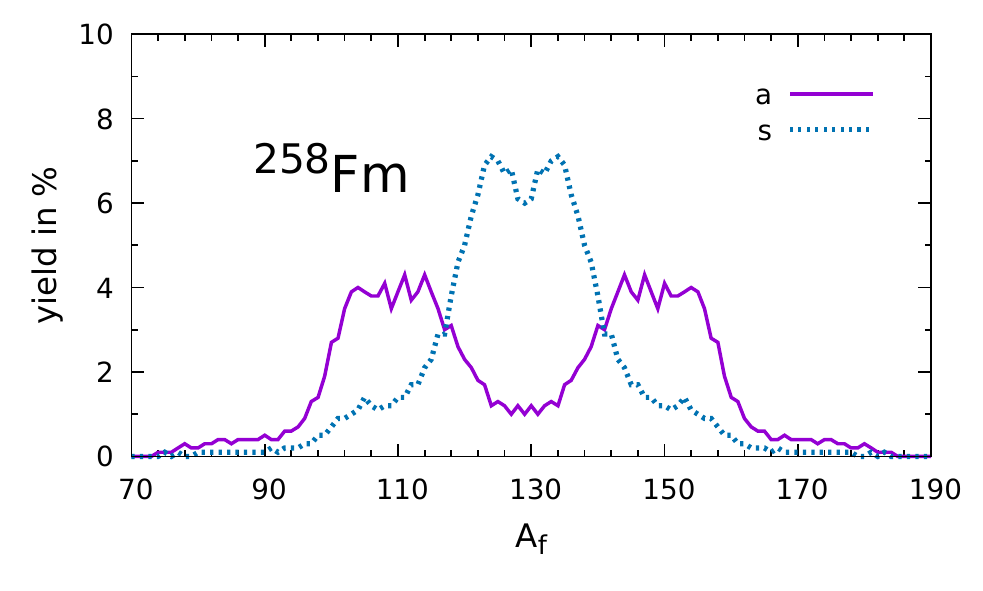}\\[-6ex]
\end{center}
\caption{Fragment mass yield calculated for spontaneous fission of $^{258}$Fm corresponding to Langevin trajectories starting from either the asymmetric ({\bf a}) or the compact symmetric ({\bf s}) turning point.}
\label{Fm258-my}
\end{figure}

\begin{figure}[htb]
\begin{center}
\includegraphics[width=0.4\textwidth]{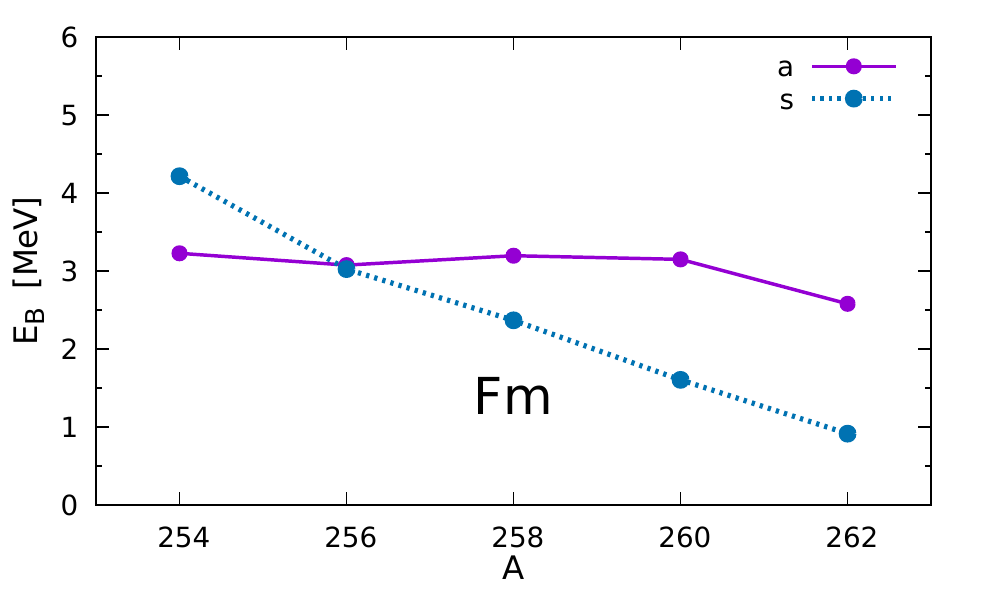}\\[-6ex]
\end{center}
\caption{Second barrier heights along the of Fm chain  corresponding to the asymmetric  {\bf a } and symmetric {\bf s}   fission paths as a function of Fm isotope  mass.}
\label{fm-e2s}
\end{figure}

The mass yields calculated for spontaneous fission of $^{258}$Fm and corresponding to the starting points {\bf a} and {\bf s} are shown separately in Fig.~\ref{Fm258-my}, being respectively asymmetric and symmetric as noted above. The final mass yield is, of course, a weighted sum of these two distributions. The weight suited for {\bf a} and {\bf s} depends  on the penetration probability ($P_i$) of the fission barrier evaluated along the path ${\cal L}_i$, which ends at the $t$-th turning-point ({\bf a} and {\bf b}). As deduced from Fig.~\ref{Fm-maps}, there are two distinct outer saddle points for the $^{256-260}$Fm isotopes, and tentatively also for $^{254}$Fm. The height of the corresponding outer barriers is plotted in Fig.~\ref{fm-e2s}. They are almost identical for $^{256}$Fm, while in the heavier isotopes, the symmetric barrier is lower than the asymmetric one.  This difference in the saddle-point heights indicates that compact symmetric fission should prevail for isotopes heavier than $^{256}$Fm. In order to calculate the final mass yield expected for spontaneous fission and compare quantitatively with experiment (wherever available), we proceed as follows.\\
The final fission fragment yield ($Y_{\rm th}$) is taken as the weighted sum of the yields $Y_a$ and $Y_s$ obtained using the points {\bf a} and {\bf s} as initial points  of the Langevin trajectories:
\begin{equation}
Y_{\rm th}(A_f)=P_a\cdot Y_a(A_f)+P_s\cdot Y_s(A_f)~,
\end{equation}
where $P_a$ and $P_s$ are the relative probabilities of reaching points {\bf a} and {\bf s} by tunneling of through fission barrier. We follow here the approximation described in Ref.~\cite{PDN22} to evaluate $P_i$.
 
\begin{figure}[t!]
\begin{center}
\includegraphics[width=0.4\textwidth]{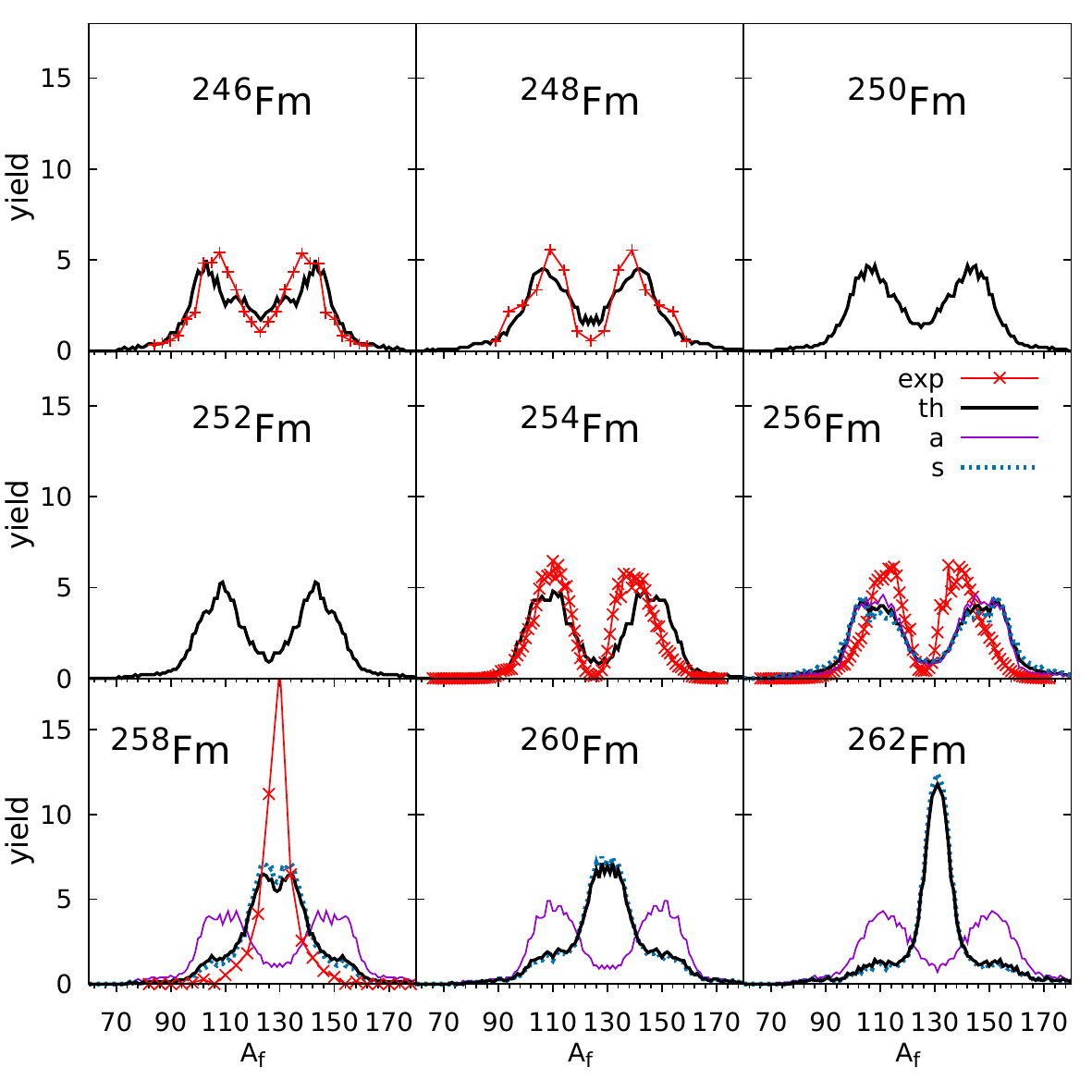}\\[-8ex]
\end{center}
\caption{Fission fragment mass yields along the Fm isotopic chain. The calculation (solid black line) is compared with experimental data for pre-neutron yields (red +) \cite{Rom10,Hof80} or post-neutron yields (red x) \cite{SJA16,Hul86} depending on availability (the little shift between pre- and post-neutron mass distributions is of no importance for the present comparison). The theoretical curves corresponding to the asymmetric (thin purple line) and symmetric (dotted blue line) fission paths are shown separately for reference, see the text.}
\label{Fm-my}
\end{figure}

In the WKB approximation, the barrier penetration probability is given by
\begin{equation}
W_i=\frac{1}{1+exp[2S({\cal L}_i)]}~,
\end{equation}
where $S({\cal L}_i)$ is the action integral taken along the ${\cal L}_i$ path
\begin{equation}
S({\cal L})= \int\limits_{s_l}^{s_r} \sqrt{{2\over \hbar^2}B_{ss}(s)
             [V(s)-E_0]}\,ds~.
\end{equation}
Here $s_l$ and $s_r$ are the left and right turning points at the path ${\cal L}$. $B_{ss}$ and $V(s)$ are the collective inertia and potential along the path ${\cal L}$ respectively, and $E_0$ is the ground state energy. The total penetration probability of the barrier is the sum of the probabilities along the asymmetric and symmetric paths. So, the relative population of the asymmetric and the compact-symmetric valley is
\begin{equation}
P_a=\frac{W_a}{W_a+W_s}~~~{\rm and}~~~P_s=\frac{W_s}{W_a+W_s}~.
\end{equation}
Following the above recipe, the final fission fragment mass yields (thick black line) predicted for spontaneous fission of the even-even $^{246-262}$Fm isotopes are shown in Fig.~\ref{Fm-my}. The yield distributions due to path {\bf a} (thin purple line) and to path {\bf s} (dotted blue line) are also displayed for reference. For the  lighter $^{254-256}$Fm isotopes, the mass yields do not depend on the choice of the starting point, while for the heavier ones, they differ significantly. One obtains the asymmetric mass yield (solid line) when starting from the point {\bf a} and the symmetric distributions corresponding to the initial point {\bf s}. The maximum of the asymmetric component in the final distribution is located between $A\approx$ 146 and 150 for the heavy fragment depending on fissioning mass, while for the symmetric component, there are two close-lying maxima with the heaviest one is sitting at A$_h=132$; the light partner is given by the fissioning mass. Comparison between the final calculation and experiment is seen to be pretty good, bearing in mind the simple recipe outlined above. Further improvement requires to consider the full dynamics of the process starting from {\it e.g.} the second minimum like in Ref.~\cite{UII19} (rather than assuming simple tunneling through each barrier separately). Work in this direction is the scope of future enhancement of the model. 

\begin{figure}[b!]
\begin{center}
\includegraphics[width=0.4\textwidth]{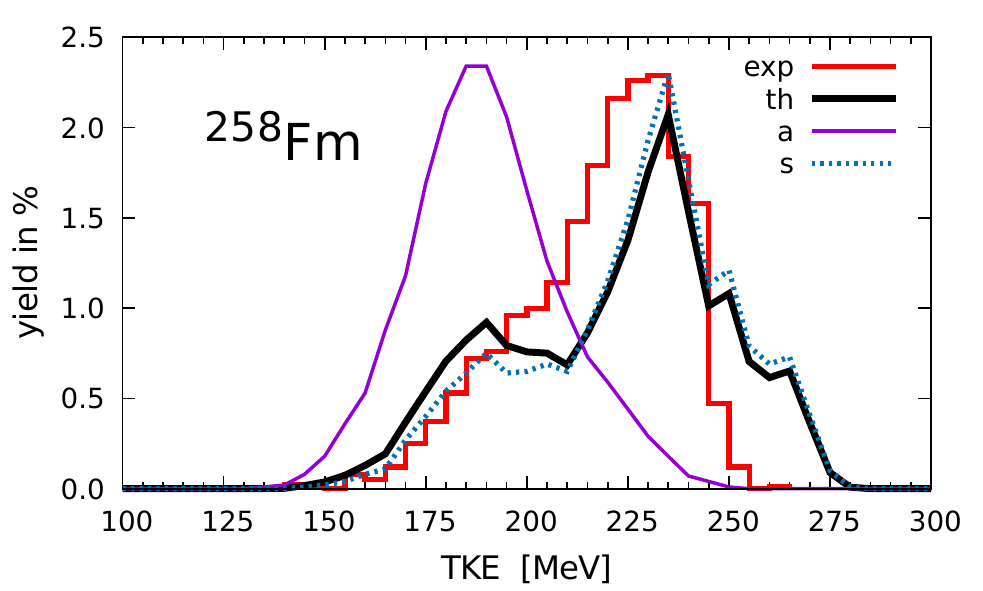}\\
\includegraphics[width=0.4\textwidth]{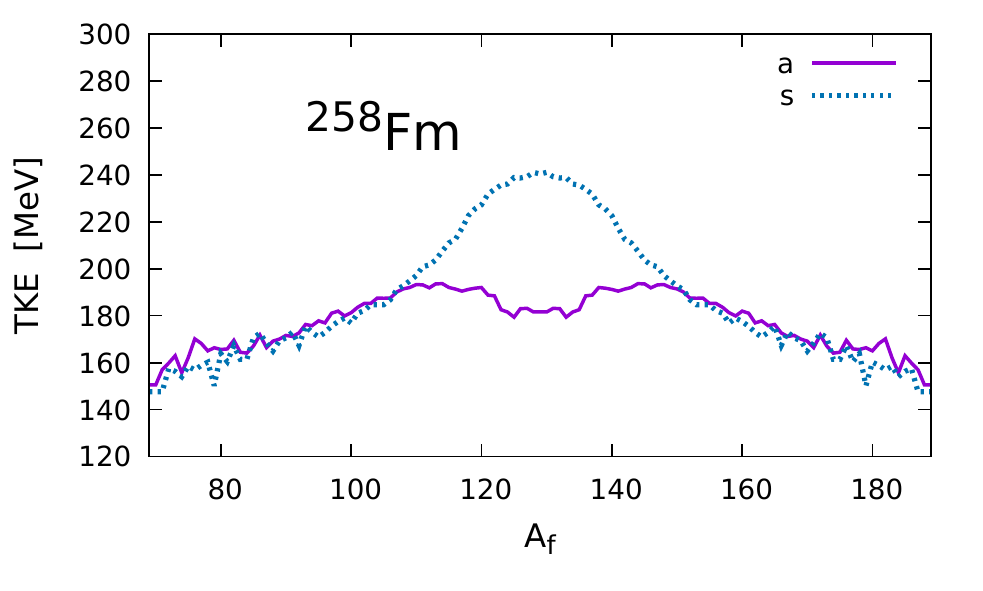}
\end{center}
\vspace{-1cm}
\caption{Top: fission fragment TKE distribution for spontaneous fission of $^{258}$Fm. The final calculated distribution is shown with the thick black full line, while the distributions characteristic of paths {\bf a}(thin purple full)  and {\bf s} (dotted blue) are displayed for reference. Experimental data (red histogram) are taken from Ref.~\cite{Hul86}. Bottom: mean TKE as a function of the fragment mass for paths {\bf a} and {\bf s}.}
\label{Fm258-spec}
\end{figure}

The calculated final fission fragment TKE distribution for spontaneous fission of $^{258}$Fm is shown in the top panel of Fig.~\ref{Fm258-spec}. This weighted TKE yield (thick black full line) can be compared with the experimental data (red histogram) \cite{Hul86}. The TKE yields corresponding to the starting points {\bf a} (thin purple) and {\bf s} (dotted blue) are drawn also. The theoretical result is seen to reproduce very reasonably the measurement. In particular it exhibits the two-humped pattern mentioned in introduction. According to the discussion above, for the $^{258}$Fm isotope, the contribution from path {\bf s} dominates. In this respect, it is important to note that Fig.~\ref{Fm258-spec} suggests that the low-energy component of the TKE distribution originates almost exclusively from path {\bf s} rather from path {\bf a}. In still other words, path {\bf s} has itself a two-humped distribution, {\it i.e.} has contributions from two different modes. This could already been seen in Figs.~\ref{Fm258-my} and \ref{Fm-my}, where some asymmetric wings appear next to the symmetric peak in the mass distribution of path {\bf s}. The mean TKE as a function of the fission fragment mass is plotted in the bottom part of Fig.~\ref{Fm258-spec}. It is seen there that for $A_f=258/2$, the TKE corresponding to the path {\bf s} is about 50\% larger than the one related to the path {\bf a}. It shows that in the case {\bf s}, for most symmetric events, one deals with a compact symmetric path. The TKE spectra confirm conclusions drawn from the mass yields comparison: in $^{258}$Fm, the compact-symmetric fission predominates.\\

\begin{figure}[h!]
\includegraphics[width=0.5\textwidth]{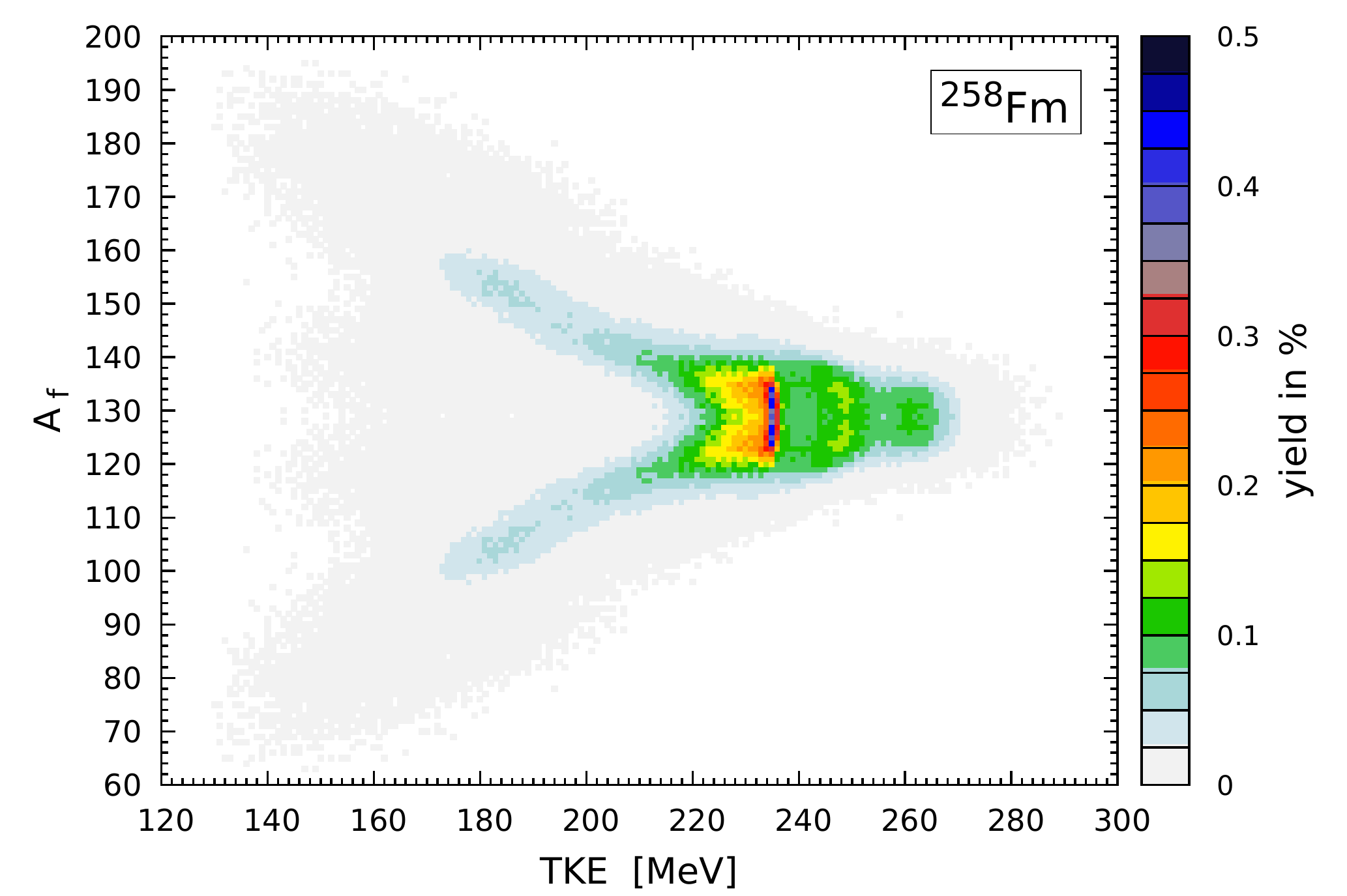}\\
\includegraphics[width=0.5\textwidth]{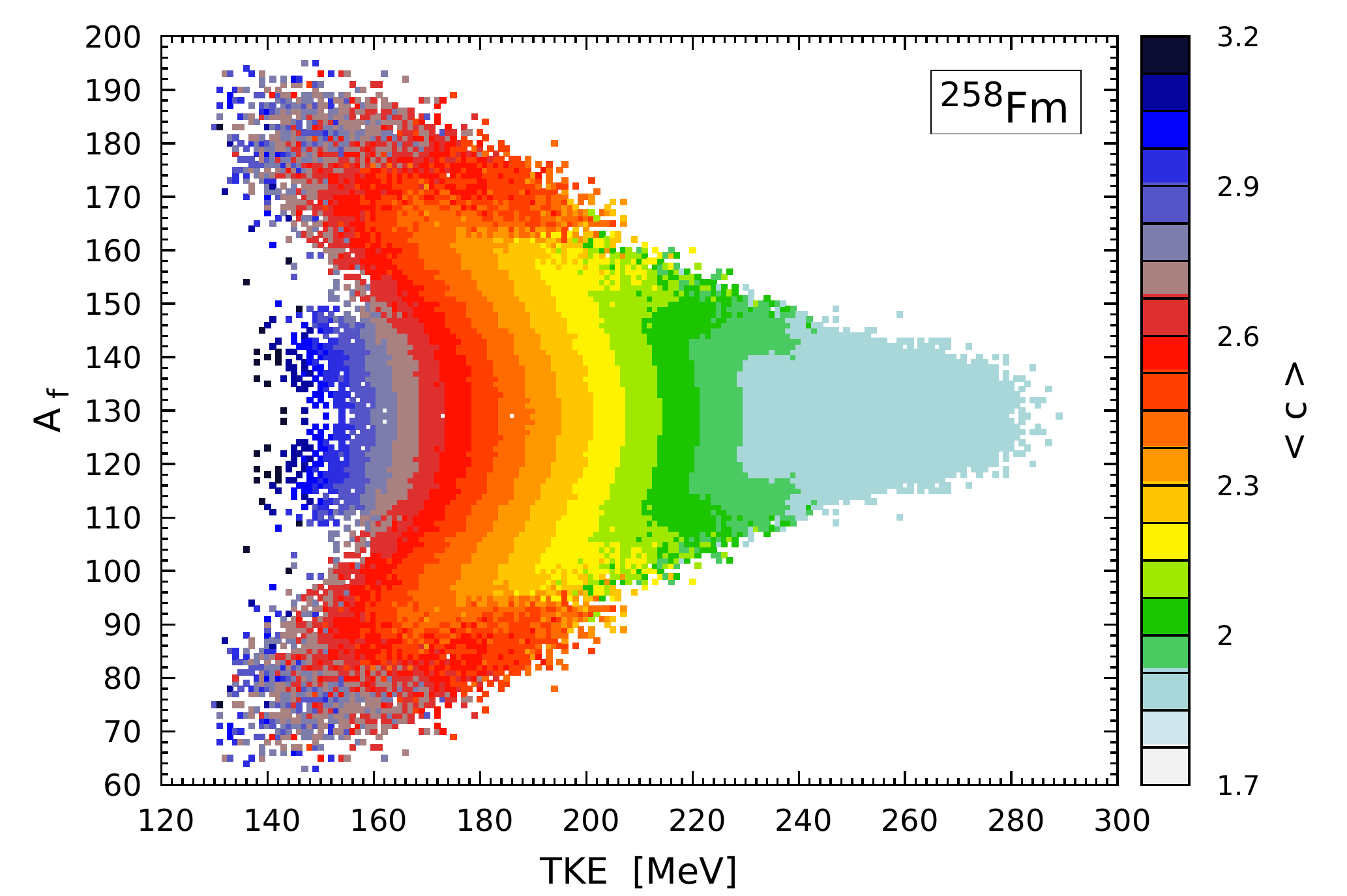}\\[-6ex]
\caption{Fission fragment yields (upper panel) and corresponding average elongation $<c>$ close to scission (bottom panel) for spontaneous fission of $^{258}$Fm as a function of fragment TKE and mass ${\rm A_f}$.} 
\label{Fm258-ymap}
\end{figure}

To get a deeper insight into the above observations and discussion, we consider higher-fold correlations. Figure ~\ref{Fm258-ymap} displays the fragment yield (top) and the elongation of the system just before scission, at the end of the trajectory (bottom) as a function of fragment mass and TKE for spontaneous fission of $^{258}$Fm. The upper panel exhibits the dominant symmetric component with TKE $\approx$ 234 MeV (dark blue and red blob) and the small contribution from the asymmetric Standard II mode at (A$_l$, A$_h$) $\approx$ (113, 145) and TKE $\approx$ 180 MeV (light blue bands), see also bottom of Fig.~\ref{Fm258-spec}. In addition, some slightly less asymmetric component is dragging from the symmetric high-TKE region down to TKE's as low as $\approx$ 129 MeV. These events correspond to the asymmetric wings of path {\bf s} mentioned above. A further insight can be obtained from the bottom panel of the Fig ~\ref{Fm258-ymap} which informs about the elongation close to scission. The dominant symmetric component originating from path {\bf s} is seen to be characterized by the smallest elongation $c \approx$ 2 at scission, confirming that it corresponds to a compact symmetric mode. The Standard II asymmetric mode has a somehow larger mean $c \approx$ 2.2-2.4 at scission as expected. But maybe most interesting is to notice that the slightly asymmetric wings dragged from symmetry to very low TKE, and which end as distinct blue blobs clearly separated from Standard II, have a mean elongation at scission above $c \approx$ 2.7, {\it viz.} larger than Standard II. That corroborates that the asymmetric wings from path {\bf s} discussed above do not originate from the Standard II due to path {\bf a}, but they shall rather to be considered as the asymmetric tails of a symmetric elongated mode stemming from path {\bf s}. It is to be noted that these tails dominate the mass distribution of path {\bf s} in $^{256}$Fm, while the symmetric compact mode prevails in $^{258}$Fm (see purple curves in the corresponding panels of Fig.~\ref{Fm-my}). This can be best understood from a detailed look at Fig.~\ref{Fm-maps}, which shows that these two "sub-paths" separate at $c \approx$ 1.7 and $a_4 \approx$ 0.54. The small differences between the PEL's of $^{256}$Fm and $^{258}$Fm drive the system in one or the other sub-path (in addition to the influence of the dynamics).

\begin{figure}[h!]
\includegraphics[width=0.5\textwidth]{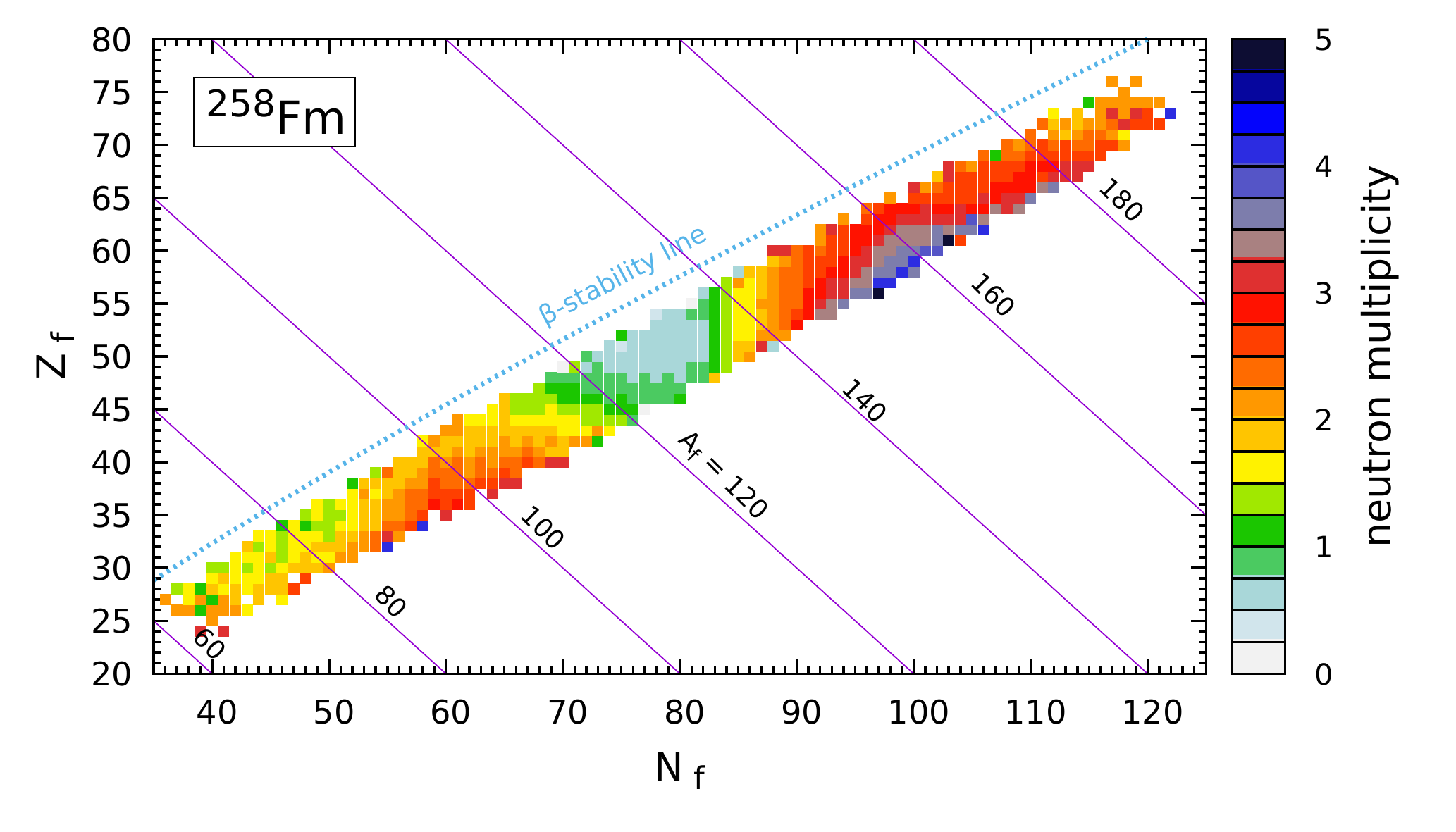}\\[-6ex]
\caption{Post-scission neutron multiplicity for spontaneous fission of $^{258}$Fm as a function of fragment neutron ${\rm N_f}$ and proton ${\rm Z_f}$ number.}
\label{Fm258-Nnz}
\end{figure}

Finally, the predicted post-scission neutron multiplicities for spontaneous fission of ${\rm ^{258}Fm}$ are shown in Fig.~\ref{Fm258-Nnz} as a function of the fragment (${\rm N_f,\,Z_f}$) isotopic composition. The diagonal purple lines correspond to constant masses. Obviously, the number of emitted neutrons at a given mass grows with the distance from the $\beta$-stability line. For the Standard II mode, the heavy and light fragments emit on the average a comparable number of neutrons, consistent with experimental observation in the region (see {\it e.g.} \cite{gindler:1979}), although evaporation may be slightly overestimated for the heavy partner (see also Fig.~\ref{nmTKE}). The compact symmetric mode exhibits the lowest post-scission multiplicity, only 0.5 neutron on the average, in line with the above discussion: the fragments of this mode are close to magic nuclei, poorly excited at scission and which experience very little shape relaxation after scission. The asymmetric wings of the mass distribution of path {\bf s} which we identified above as stemming from an elongated symmetric mode show post-scission multiplicity values which are somehow intermediate between those of Standard II and of the compact symmetric mode, while it would be expected that these events exhibit the largest post-scission multiplicities. Overall the model thus describes the main trends observed in experiment in the region. The deficiency regarding a more detailed quantitative description is mostly due to the limitation of the theory in terms of full variety of shapes, in particular at scission, of the energy sharing prescription already mentioned above.

\section{Conclusions}

We have proposed the innovative Fourier over Spheroid (FoS) prescription as a fast and flexible nuclear shape parametrization to model fission by means of four collective coordinates {\it i.e.} elongation, left-right asymmetry, neck size, and non-axiality. Neglecting non-axiality from the outer saddle region to scission, we have developed a new 3D Langevin code, based on the FoS, the LSD + Yukawa folded macroscopic-microscopic potential energy landscape, a procedure to account for charge equilibration at scission, and a method to compute the excitation energy available in the primary fragments. Finally, the de-excitation of the latter after scission was computed. Altogether gives access to a wide palette of observables, treated in a consistent way, and which permits to analyze high-fold correlations. Such information is crucial to evaluate in a unambiguous way the reliability of specific theoretical prescriptions which are often entangled in the intricate fission process.\\
The model was first tested and tuned to reproduce at best experimental observation from thermal neutron-induced fission of $^{235}$U. In a second step, it was applied to fission along the Fm isotopic chain, and seen to explain the famous abrupt transition observed in the fragment properties between $^{256}$Fm and  $^{258}$Fm.\\
The achievement of the present model is estimated to be impressive considering its relative simplicity. Remaining discrepancies are ascribed to limitations mainly in terms of dimensionality of the shape parametrization, restriction to the outer-saddle to scission dynamics, charge equilibration and energy sharing recipes at scission, and possibly the neglect of angular momentum. Work to improve along these lines is foreseen. Also, the extension of the model to account for multi-chance fission is underway. This enhancement is very important for calculations of interest in nuclear energy applications. Further calculations for wider mass and excitation energy ranges of the fissioning nucleus, and comparison with experiment wherever available, are in progress in parallel. These are important to constrain more and more the model ingredients, and refine them. Independent of these developments, the model constitutes already a useful tool for various domains where systematic and fast predictions are required. Additionally, the conclusions drawn from its comparison with experiment can be a useful guidance for more fundamental theory.
\\
\\
{\bf Acknowledgments}\\

We acknowledge discussions with F. A. Ivanyuk. The authors would like to thank A. G\"o\"ok and A. Al-Adili for supplying us with some experimental data. This work has been supported by the Polish National Science Center (Grant No. 2018/30/Q/ST2/00185) and by the Natural Science Foundation of China (Grant No. 11961131010 and 11790325).\\[4ex]


\setcounter{equation}{0} 
\renewcommand{\theequation}{A-\arabic{equation}}

\noindent
{\bf{\large Apendix:} Deformation of fission fragments}\\
 
At the end of each thousand Langevin trajectories, i.e., at scission configuration, one has to determine the deformation parameters of both fission fragments. This procedure has to be repeated several thousand times, so it should be rapid. Knowledge of the fragment deformations is necessary to estimate its deformation energy, which contributes significantly to the fragment excitation energy (\ref{Eteq}).
\begin{figure}[h!]
\includegraphics[width=0.5\textwidth]{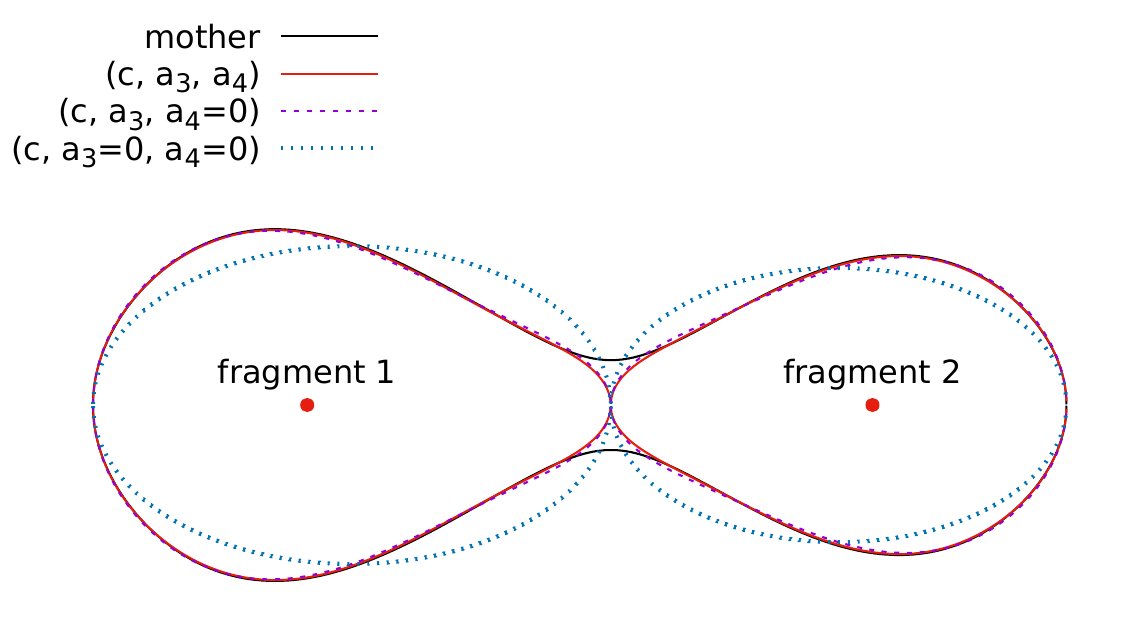}\\[-6ex]
\caption{Shape of the mother nucleus (black line) and the fragments fitted using only the elongation parameter $c$ (dotted line), $c$ and the pear-like deformation $a_3$ (dashed line), and three deformation parameters $c,\,a_3$, and $a_4$ (red line).}
\label{Ffit}
\end{figure}

 Let us assume the fission fragments have the masses $A_1$ and $A_2$, where $A=A_1+A_2$ is the mass of the mother nucleus described by the profile $\rho^2(z)$ Eq.~(\ref{rhos}). The following data on the mother  nucleus around scission are needed to determine the deformation of the fragments:
\\[-3ex]
$$z_{\rm min}=-z_0+z_{\rm sh},~~~~~~~z_{\rm neck},~~~~~~~~
z_{\rm max}=z_0+z_{\rm sh},$$
$$z_{\rm cm}(1),~~z_{\rm cm}(2),~~\rho^2({z_{\rm min}+z_{\rm neck}\over 2}),
~~\rho^2({z_{\rm neck}+z_{\rm max}\over 2}).$$
The corresponding spherical radii are $R_{01}$, $R_{02}$, and $R_0$, where $R_{0i}=R_0(A_i/A)^{1/3}$. The fragment elongations are: 
\begin{equation}
\begin{array}{ll}
c_1=&{\displaystyle {z_{\rm neck}-z_{\rm min}\over 2R_{01}}~,}\\[+2ex]
c_2=&{\displaystyle {z_{\rm max}-z_{\rm neck}\over 2R_{02}}~.}
\end{array}
\end{equation}
One evaluates the reflection asymmetry parameter $a_{3i}$ from the shift of the fragment mass center with respect to its geometrical center:
\begin{equation}
\begin{array}{ll}
a_{31}=&{\displaystyle -{2\pi\over 3c_1R_{01}}\left[z_{\rm cm}(1)-{z_{\rm neck}-z_0+z_{\rm sh}\over 2}\right]~,}
\\[+2ex]
a_{32}=&{\displaystyle -{2\pi\over 3c_2R_{02}}\left[z_{\rm cm}(2)-{z_0+z_{\rm sh}-z_{\rm neck}\over 2}\right]~.}
\end{array}
\end{equation}
To determine the $a_{4i}$ deformation of the fragment $i$ one uses the FoS relation:
\begin{equation}
\rho^2_i(0)={R_{0i}^2\over c_i}f(0)={R_{0i}^2\over c_i}
\left(1-{4\over 3}a_{4i}-{4\over 5} a_{6i}\dots\right)\,,
\end{equation}
where 
\begin{equation}
\begin{array}{ll}
\rho^2_1(0)=&{\displaystyle \rho^2[(z_{\rm neck}+z_{\rm max})/2]~,}\\[+3ex]
\rho^2_2(0)=&{\displaystyle \rho^2[(z_{\rm neck}+z_{\rm min})/2]~.}
\end{array}
\end{equation}
Assuming that $a_6=-a_4/10$ (LD energy minimum) one obtains:
\begin{equation}
\rho^2_i(0)={R_{0i}^2\over c_i}\left(1-{96\over 75}a_{4i}\right)
\end{equation}
what implies
\begin{equation}
a_{4i}={75\over 94}\left(1-{c_i\over R_{0i}^2}\,\rho^2_i(0)\right)~.
\end{equation}
The quality of the shape described above is shown in Fig.~\ref{Ffit}. It is seen that taking into account the pear-like deformation significantly improves the quality of the fit, while the $a_4$ deformation has only a tiny effect. 
                                  


\end{document}